\RequirePackage{fix-cm}
\documentclass[smallcondensed]{svjour3}     % onecolumn (ditto)
\smartqed  % flush right qed marks, e.g. at end of proof
\usepackage{graphicx}
\usepackage{algorithm}
\usepackage{amssymb}
\usepackage[noend]{algpseudocode}
\usepackage[multiple]{footmisc}
\usepackage{epsfig}

\begin{document}

\title{Genetic algorithm for multi-objective optimization of container allocation in cloud architectures%\thanks{Grants or other notes
}

\titlerunning{GA for multi-objective optimization of container allocation in cloud}        % if too long for running head

\author{Carlos Guerrero         \and
        Isaac Lera \and Carlos Juiz %etc.
}


\institute{C. Guerrero (Corresponding author), I. Lera, C. Juiz \at
              Computer Science Department, University of Balearic Islands, Crta. Valldemossa km 7.5, E07122 Palma, Spain \\
              Tel.: +34-971172965\\
              Fax: +34-971173003\\
              C. Guerrero\\
              \email{carlos.guerrero@uib.es} \\
              I. Lera\\
              \email{isaac.lera@uib.es} \\
              C. Juiz\\
              \email{cjuiz@uib.es}           %  \\
 %          S. Author \at
               .          %  \\
}

\date{Received: date / Accepted: date}

\maketitle

\begin{abstract}
	The use of containers in cloud architectures has become widespread because of advantages such as limited overhead, easier and faster deployment and higher portability. Moreover, they are a suitable architectural solution for deployment of applications created using a microservices development pattern. Despite the large number of solutions and implementations, open issues have not been addressed in container automation and management. Container resource allocation influences system performance and resource consumption so it is a key factor for cloud providers. We propose a genetic algorithm approach, using the Non-dominated Sorting Genetic Algorithm-II (NSGA-II), to optimize container allocation and elasticity management due to the good results obtained with this algorithm in other resource management optimization problems in cloud architectures. The optimization has been focused on a tight use of the resources and a reduction of the network overhead and system failure rate. A model for cloud cluster, containers, microservices and four optimization objectives is presented. Experimental results have shown that our approach is a suitable solution to address the problem of container allocation and elasticity and it obtains better objectives values than the container management policies implemented in Kubernetes.

\end{abstract}

\section{Introduction}

Microservices architecture is a development pattern that has grown in popularity in recent years. This pattern defines an application as a set of independent small and modular services executing a single task and the specification of the interoperability of these microservices to achieve the application requirements. Several companies have adopted this practice and traditional service cloud applications are being migrated to microservices~\cite{balalaie16microservices} for all their benefits: making easier to ship and update applications, allowing independently update and redeploy of parts of the application, getting closer development and operations teams, allowing continuous release cycle, simplifying orchestration of applications across heterogeneous cloud data centers and so on~\cite{fazio16open}.

The containerization of applications is one of the enabling technologies for microservices architectures. Containerization is a method for operating system (OS) virtualization. Virtual instances of OS, called containers, are  a form of isolation for the OS environment and its file system, and they are executed over a single host and a single kernel. Therefore, each microservice of an application is deployed in a container, without launching a complete virtual machine. By this, system overhead is reduced as security and isolation of the environments are maintained. The deployment of a microservice is as simple as starting the execution of a new container of the microservice. Therefore, microservices can be scaled up by just creating new containers until the desired scalability level is achieved. Docker is one of the most successful implementations of container architectures~\cite{peinl2016docker}. Container clusters have been developed to easily deploy containers in the cloud and to enable features as consolidation, elasticity, load balancing and scalability in container architectures. Docker Swarm, Apache Mesos and Google Kubernetes are examples that provide automatic deployment and orchestration~\cite{peinl2016docker}.

Despite the fast growth of these technologies and the existence of \textit{de-facto} solutions, there are important open issues in scheduling and management of microservices in the cloud. Allocation of containers across the physical nodes of a cluster influences system performance and reliability, as well as elasticity, and its automatic management, influences application performance. The problem of elasticity in virtual machines environments has been widely studied~\cite{caballer2015dynamic} but, to the best of our knowledge, not in the case of container-based clouds. Current container cluster managers, as Kubernetes, implement simple approaches for the automatic scalability of containers and the allocation of containers to physical machines, only focused on physical resource usages and general threshold values~\cite{fazio16open}. We propose to improve the elasticity schedulers and runtime container allocators to obtain better results in terms of system provisioning, system performance, reliability and network overhead, four of the traditional research challenges in cloud architectures~\cite{zhang2010cloud}.

%Resource management and efficient allocation help to handle the fluctuations in cloud workloads to guarantee QoS (Quality of Service) and provisioning~\cite{jennings2015resource}. The physical resources are limited and the architecture has to ensure their efficient and fair share among users. 

Resource management optimization is a NP-complete problem and it needs to be addressed by metaheuristic approaches~\cite{wei2010game}. The optimal solution can be only obtained with an evaluation of all the possible combinations. The literature reflects that evolutionary approaches, as Genetic Algorithms (GA), are common solutions to resource management in cloud~\cite{zhan2015cloud}. Furthermore, Non-dominated Sorting Genetic Algorithm-II (NSGA-II) is one of the most usual solutions for multi-objective optimization~\cite{vonlucken2014survey}.

We propose to use NSGA-II to impolement the container allocation strategy and the automatic elasticity management (scalability level of the microservices) by the optimization of four objectives:
\begin{itemize}
	\item the provisioning of new applications and the elasticity of the currently deployed ones by keeping a uniform distribution of the workload across the cluster; 
	\item the performance of the deployed applications and their assigned resources by considering a suitable scalability level of their microservices and a uniform distribution of the microservices workload across their containers; 
	\item the reliability of the microservices by considering a suitable scalability level and a correct distribution avoiding single points of failure;
	\item the network overhead of the communications between microservices by placing the containers of related microservices in physical machines with short network distances.
\end{itemize}

We summarize the main contributions of our work in:
\begin{itemize}  
	\item The improvement of resource management in container clusters by optimizing container allocation to physical machines and the automatic container scalability;
	\item The use of four decision criteria ---microservices reliability, microservices network overhead, balanced use of the physical machines and balanced workload over the containers--- for the resource management optimization;
	\item The implementation and the study of an evolutionary approach, the genetic algorithm NSGA-II, to solve this type of multi-objective optimization problem;
	\item The validation of our approach with workload and system models based on data from real deployed scenarios .
\end{itemize}
  
\section{Related work}

Resource management optimization is a common research topic in the field of Cloud Computing~\cite{manvi2014resource}. Several surveys have been published including a wide range of techniques to address resource management problems as, for example, scheduling~\cite{singh2016survey}, provisioning~\cite{sukhpal2016cloud} and allocation~\cite{beloglazov2012energy,akhter2016energy}.

In addition, evolutionary approaches and genetic algorithms are commonly used in resource management in cloud environments. The adequacy of genetic algorithms in general, and NSGA-II in particular, has been widely demonstrated along the research bibliography~\cite{zhan2015cloud}. Most of the researches are addressed to improve the placement of virtual machines in physical machines based on objectives from very different points of view, e.g. performance, reliability or energy efficiency. For example, Pascual et al. studied three multi-objective optimization algorithms, including NSGA-II,  and demonstrated that the use of evolutionary approaches is a low-cost optimization solution to obtain better performance and to reduce the energy consumption~\cite{pascual2015towards}.  

There is a wide range of implementations of container-based solutions but issues as container orchestration and self-management remain open~\cite{peinl2016docker}. The number of studies for resource management in container-based architectures is very limited. To the best of our knowledge, our solution is the first approach that uses genetic algorithms and only three recent works deals with container allocation problem: two of them based on Linear Programming Solvers and another one based on game theoretic problem.

Guan et al.~\cite{guan2016application} proposed an application-oriented algorithm that decides the physical node where a new container is scheduled by minimizing the application deployment cost. Kang et al. also implemented a container scheduler for new arrivals of containers but considering the objectives of saving energy consumption and guaranteeing the performance level~\cite{kang2016cloud}. These two first works differs from our proposal because they solve the allocation problem only for new container arrivals while the current allocated containers are not migrated between machines. 

Finally, Xu et al. presented an approach based on the Stable Marriage Problem, a many-to-one matching problem, to reduce the response time of costumers' jobs and to improve the physical computational resources utilization~\cite{xu2014novel}. In this approach, not only the optimization objectives, , or satisfaction as they called it, need to be measured  but also a rank of physical machines for each container and a rank of containers for each physical machine need to be calculated. These rank functions are based on criteria as selecting the machine with more resources, or the proportionality between container resource consumption and machine capacity. When a physical machine and a container are the best ranked between themselves, the container is allocated in the machine. The satisfaction function is used after each new matching to ensure that the allocation improves the optimization objectives. The allocation of the machines is driven by the rank functions instead of the optimization objectives. In our approach, rank functions are not considered, and our objective functions can not be directly translated to rank functions.

Our proposal is to use NSGA-II, a general accepted approach for multi-objective optimization problems in cloud architectures, in a resource management problem, container allocation and elasticity that, from our current knowledge, has not been previously addressed with evolutionary strategies.

\section{System model}
\label{systemmodel}

Our approach needs to use a formal definition of the system by modeling applications, microservices, physical machines and network. For a better understanding, Table~\ref{systemmodelsummary} summarizes the parameters of the system model described in the remainder of this section.

We consider a set of applications $A$ following a development pattern based on microservices. Each of these applications ($app_j$) is characterized with the number of user requests ($ureq_j$) and the microservices stack. The microsevices stack is the set of microservices and the interoperability between them that implements the requirements of an application. These interoperability relationships are established when microservices consume results generated by other microservices. By this, the microservices stack is modeled as a directed graph. The nodes are the microservices ($ms_i$) and the edges represent the interoperability between them. Two nodes are connected by an edge, $(ms_{provider},ms_{consumer})_{prov/cons}$, if the target microservice $ms_{consumer}$ consumes the results of the origin one $ms_{provider}$. 

Each microservice is characterized as a tuple $(msreq_i, res_i, thr_i, fail_i)$ where, $msreq_i$ is the number of microservice requests needed to satisfied one application user request ($ureq_j$); $res_i$ is the computational resources consumed to satisfied one request of the microservice; $thr_i$ is the threshold level for the resources consumption above which the service performance is degraded and the microservice results on a bottleneck for the application $app_j$; and $fail_i$ is the failure rate of the microservice. The values for $ureq_j$ and $msreq_i$ depend on the workload of the system and the values for $res_i$, $thr_i$ and $fail_i$ depend on the implementation of the microservices.

Each microservice is executed in the system encapsulated in one or more containers $cont_k$, represented as $ms_i \equiv cont_k$. The number of containers depends on the scale level of the microservice ($scale_i$). The microservice resource consumption is divided uniformly among the containers, so the computational resource consumption of the container $res_k$  is calculated as $\frac{ureq_j \times msreq_i \times res_i}{scale_i}$.

The set of containers ($C$) is executed in a container cluster with a set of physical machines $pm_l$. The relationship $alloc(cont_k)=pm_l$ indicates that container $cont_k$ is allocated to the physical machine $pm_l$. The notation can be summarized to represent the physical machine that allocates a container of a microservice $ms_i$ using $alloc(ms_i)=pm_l$ if $cont_k \equiv ms_i$ and $alloc(cont_k)=pm_l$.

Each physical machine is characterized by the tuple $(cap_l, fail_l)$ where $cap_l$ is the computational capacity and $fail_l$ is the failure rate of the node. A constraint of the system is that the sum of computational resources used by the containers allocated to a physical machine needs to be lower than the computational capacity of this physical node. Finally, the physical machines are interconnected with a physical network $N$, where the paths between nodes $pm_{l}, pm_{l'}$ are characterized by their network distance $dist_{pm_{l},pm_{l'}}$.

\begin{table}
	\caption{Summary of the system model}
	\label{systemmodelsummary}       % Give a unique label
	\begin{tabular}{lll}
		\hline\noalign{\smallskip}
		Element & Parameter & Description  \\
		\noalign{\smallskip}\hline\noalign{\smallskip}
		Application   & $app_j \in A$ & application with id. $j$\\
		& $|A|$ & total number of applications in the cluster \\
		& $ureq_j$ & number of user request for application $app_j$ \\
		Microservice& $ms_i \ in S$ & microservice with id. $i$\\
		& $|S|$ & total number of microservices \\
		& $(ms_{i'},ms_i)_{prod/cons}$ &  microservice $ms_i$ consumes microservice $ms_{i'}$\\
		& $msreq_i$ & microservice requests number $ms_i$ \\
		&&needed for each $ureq_j$ request\\
		&& from application $app_j$\\	
		& $res_i$ & computational resources required for a\\
		&& microservice request\\
		& $thr_i$ & threshold for normal operation of a\\
		&& microservice $ms_i$\\
		& $fail_i$ & failure rate for a microservice $ms_i$\\
		Container      & $cont_k \in C$ & container with id. $k$\\
		& $|C|$ & total number of containers \\
		& $cont_k \equiv ms_i$ & container $cont_k$ encapsulate/execute\\
		&& microservice $ms_i$\\
		& $scale_i$ & number of containers for $ms_i$\\
		& $res_k$ & computational resource consumption of\\
		&&container $cont_k$\\
		Physical machine      & $pm_l \in P$ & physical machine with id. $l$\\
		& $|P|$ & total number of physical machines \\
		& $alloc(ms_i / cont_k) = pm_l$& physical machine $pm_l$ allocates  \\
		&&service $ms_i$ / container $cont_k$\\
		& $cap_l$ & computational capacity of physical\\
		&&machine $pm_l$\\
		& $fail_l$ & failure rate for a physical\\
		&&machine $pm_l$\\
		Network       & $dist_{pm_{l},pm_{l'}}$ & network distance between $pm_{l}$ and $pm_{l2'}$\\
		
		\noalign{\smallskip}\hline
	\end{tabular}
\end{table}

\subsection{Optimization objectives}

In our approach, the optimization of the objectives is achieved through the allocation of the containers and the definition of the scale level, i.e., relationship $allocation(cont_k)$ and $scale_i$ are determined $\forall cont_k$ and $\forall ms_i$ using a set of optimization objectives. The objectives to optimize are: (OBJ-i) the container workload avoiding the presence of bottlenecks; (OBJ-ii) the balanced used of the cluster to facilitate the future admission and provision of new applications; (OBJ-iii) the reliability of the applications by an evenly distribution of the containers through the cluster nodes; and (OBJ-iv) the intercommunication overhead by allocating related microservices to physical machines with short network distances.

For the first optimization objective (OBJ-i), we consider that the workload of the containers is appropriated when the resource consumption of the container $res_k$ is similar to the resource consumption threshold of the microservice $thr_i$. Thus, a high scalability level of a microservice is penalized when the container is underused --the container resource consumption is lower than the microservice resource threshold-- and a low scalability level is penalized when the number of microservice requests is very high --the resource consumption is higher than the microservice resource threshold--. We define a metric call Threshold Distance that is the difference between the resource consumption of a container and the threshold value of a microservice. This is formalized in Equation~\ref{containerworkload}.

\begin{equation}
\label{containerworkload}
Threshold\ Distance\ = \sum_{\forall ms_i}  \left| \frac{ureq_j \times msreq_i \times res_i}{scale_i} - thr_i \right| 
\end{equation}

For the second optimization objective (OBJ-ii), we consider that the cluster is balanced when the use of the physical nodes is uniform across the cluster ---the computational resources consumption is uniformly distributed among the physical machines---. If a physical machine does not have any container allocated, it is not considered in this calculation, because it could be switched off. We propose to use the standard deviation of the percentage of resource usages of the physical nodes to evaluate the balance of the cluster~\cite{xu2016location}. This is formalized in Equation~\ref{clusterbalanced}.

\begin{eqnarray}
\label{clusterbalanced}
Cluster\ Balanced\ Used\ = \sigma(PM_{usage}^{pm_l},\quad  if\quad  \exists\ ms_i\quad |\ alloc(ms_i)=pm_l)
\end{eqnarray}
where:
\begin{eqnarray}
\label{clusterbalanced2}
PM_{usage}^{pm_l} = \frac{\sum_{ms_i}\frac{ureq_j \times msreq_i \times res_i}{scale_i}}{cap_l}\quad   \forall ms_i\quad |\ alloc(ms_i)=pm_l
\end{eqnarray}

For the third optimization objective (OBJ-iii), we measure the reliability of the system using the failure rate of the applications. An application fails when any of their microservices fails. A microservice fails when all the container replicas fail. A container fail is generated by a fail in the container, $fail_i$, or by a fail in the physical machine that allocates the container, $fail_l$. This is mathematically modeled in Equation~\ref{failure}.

\begin{eqnarray}
\label{failure}
System\ Failure\ =  \sum_{\forall ms_i}Service\ Failure (ms_i)
\end{eqnarray}
where:
\begin{eqnarray}
\label{failure2}
Service\ Failure(ms_i)\ = \\ = \prod_{\forall pm_l\ |\ allocation(ms_i)=pm_l} \left( fail_l + \prod_{\forall ms_i\ |\ allocation(ms_i)= pm_l} fail_i \right)
\end{eqnarray}

For the fourth optimization objective (OBJ-iv), we approximate the network overhead between the microservices of an application with the average network distance between all the pairs of consumer and provider containers. Consequently, we consider the mean value of the distances between the replicas of a microservice and the containers executing microservices consumed by the former. Equation~\ref{distance} formalizes this metric.

\begin{eqnarray}
\label{distance}
Total\ Network\ Distance = \sum_{\forall ms_i}Service\ Mean\ Distance(ms_i)
\end{eqnarray}
where:
\begin{eqnarray}
\label{distance2}
Service\ Mean\ Distance(ms_i) = \\ =\frac{\sum_{\forall cont_k | cont_k \equiv ms_i} \left(\sum_{\forall cont_{k'} \equiv ms_{i'}\ | (ms_{i'},ms_{i})_{prov/cons}} dist_{alloc(cont_k),alloc(cont_{k'})}\right)}{|cont_{k}| \times |cont_{k'}|}
\end{eqnarray}

To sum up, our aim can be defined as a four-objectives optimization problem where the solution establishes (a) the scalability level of each microservice and (b) the allocation of containers to the physical machines, by minimizing (OBJ-i) Threshold Distance, (OBJ-ii) Cluster Balanced Used, (OBJ-iii) System Failure and (OBJ-iv) Total Network Distance functions. The problem can therefore be formulated as:

Determine: 
\begin{eqnarray}
scale_i = |\{cont_k\}|\quad |\quad cont_k \equiv ms_i\quad \forall ms_i \\
alloc(ms_i)\quad \forall ms_i 
\end{eqnarray}

By minimizing:

\begin{eqnarray}
\sum_{\forall ms_i} \left| \frac{msreq_i \times res_i}{scale_i} - thr_i \right| \\ \sigma(PM_{usage}) \\
\sum_{\forall ms_i}Service\ Failure (ms_i) \\
\sum_{\forall ms_i}Service\ Mean\ Distance(ms_i)
\end{eqnarray}

Subject to the constraint:

\begin{eqnarray}
\label{constraint}
\sum_{\forall cont_k | alloc(cont_k) = pm_l} res_k < cap_l\quad  \forall\ pm_l
\end{eqnarray}

This optimization problem is a NP-complete problem because all the possible scalability levels and allocations should be evaluated to find the optimal solution. A metaheuristic method needs to be used to address the problem~\cite{wei2010game}.

\section{Genetic algorithm proposal}

Genetic algorithms (GA) are metaheuristic approaches for solving multi-objective optimization problems. Our solution is based on the Non-dominated Sorting Genetic Algorithm-II (NSGA-II) that is considered a standard approach for multi-objective problems. In the field of GA, and along this manuscript, the following terms are used interchangeably:  solution, chromosome and individual; set of solutions and population; algorithm iteration and generation; offspring, next population and next algorithm iteration.

GA implementation requires to define the representation of the chromosome, the crossover and mutation operators, the fitness function, the selection operator, the offspring generation and the execution parametrization. The quality of the solutions is influenced by how these elements are defined~\cite{mitchell1998introduction}. The next subsections explain the details of the adaptation of each part of the GA to our optimization problem. Additionally, we include a general structure of the genetic algorithm in Algorithm~\ref{algorithm} for a better understanding.

\subsection{Chromosome representation}

\begin{figure*}
	\includegraphics[width=0.35\textwidth]{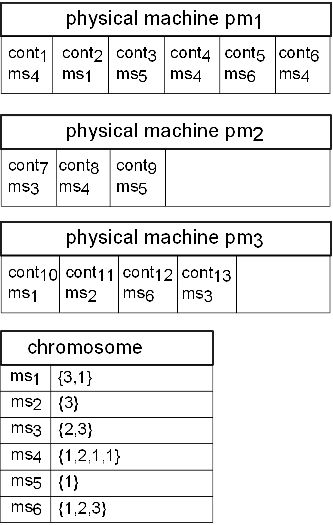}
	\caption{Chromosome representation example for a system with 3 physical machines allocating 6 microservices deployed in 13 containers}
	\label{fig:chromosome}       % Give a unique label
\end{figure*}

In GA, the solutions for the problem are usually represented using a string-based notation. Each of these solutions is called chromosome. We define a string-based notation to represent the number of containers (scalability level) for each microservice, and the allocation of these containers to the physical machines (Figure~\ref{fig:chromosome}).

The representation of the allocation is an array, called microservices array, where the index is the microservice identifier and the content is a list, called allocation list, with the identifiers of the physical machines where the containers for the given microservice are allocated. The number of elements in the allocation list is the scalability level for a microservice. The same physical machine identifier can be repeated several times in the allocation list because the physical machine can allocate several containers of the same microservice.

\subsection{Crossover and mutation operators}

Each iteration of a GA generates a new population for the given problem. Each of these new generations tends to approach to better values for the objectives. This trend is based on the biological evolution concepts of crossover and mutation, where new individuals are created by mating the best ones of the previous generations. In GA, crossover and mutation operators define the way that parent solutions are combined and modified to generate new ones.

Crossover operator combines two solutions to generate, usually, two new ones. This combination is defined over the chromosome representation taking alternatively pieces from the first and the second chromosome. In our case, the crossover is applied independently to each single allocation list. In each of those lists, a single-point crossover operator is applied~\cite{mitchell1998introduction}. This operator combines two chromosome pieces, one from each parent, to generate the first individual. The second one is generated by the combination of the opposite pieces of the chromosomes.

\begin{figure*}
	\includegraphics[width=0.75\textwidth]{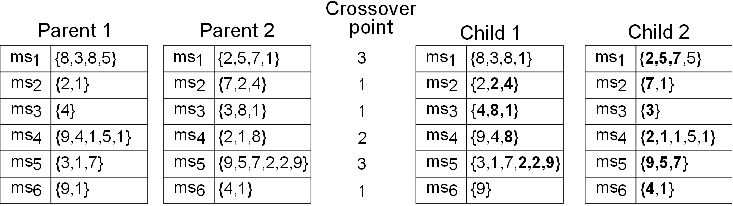}
	\caption{Example of the crossover operator applied to two parent solutions to obtain two new individuals}
	\label{fig:crossover}       % Give a unique label
\end{figure*}

The single-point crossover operator generates a random number, $r$, between 1 and the number of elements in the smaller parent allocation list, $min($$length(c1),$ $ length(c2))$. The outputs of the operator are two new allocation lists obtained from the concatenation of the $[1,r]$ elements of the first list and the $[r+1, length(c2)]$ elements of the second list ---for the first output--- and the concatenation of the $[1,r]$ elements of the second list and the $[r+1, length(c1)]$ elements of the first list ---for the second output---. A different random value $r$ is generated for each allocation list. Figure~\ref{fig:crossover} shows an example of the crossover operator applied to two parents.

Mutation operator is usually applied to just one individual. The objective of mutation operator is to modify solutions randomly to avoid local solutions in the solution search space~\cite{mitchell1998introduction} . 

\begin{figure}
	\includegraphics[width=0.35\textwidth]{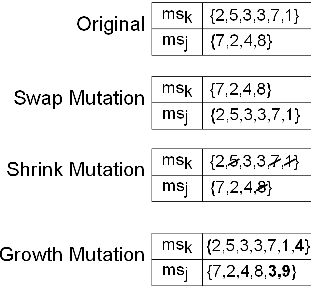}
	\caption{Examples for the three mutation operators}
	\label{fig:mutation}       % Give a unique label
\end{figure}

We have defined three mutation operators: Swap Mutation, which shuffles the array positions (allocation lists) of the microservice array; Shrink Mutation, which removes a random number of physical machines from all the allocation lists, i.e., the scalability level is decreased for all the microservices; and Growth Mutation, which increases the allocation lists by adding a random number of physical machine identifiers, i.e., the scalability level is increased for all the microservices. Figure~\ref{fig:mutation} represents examples of the three mutation operators.

\subsection{Fitness function and selection operator}

The NSGA-II algorithm includes a specific definition of how the fitness function and the selection operator have to be used. NSGA-II is designed for multi-objective optimizations. Thus, the solution is not evaluated with a single fitness value. The suitability of the solution is expressed by a tuple with the values for each of the optimization objectives. In our case, the fitness value is represented with the values of the four functions to minimize (Equations~\ref{containerworkload}, \ref{clusterbalanced}, \ref{failure} and~\ref{distance}). If the constraints for a solution are not accomplished (Equation~\ref{constraint}), the fitness value is fixed to infinite for all the criteria.

The selection operator establishes how the parents of a crossover operation are selected from the population. To explain the details of the selection operator defined in the NSGA-II, we need to previously introduce three concepts~\cite{deb2002fast}. The first one is the dominance. A solution $s_1$ dominates another solution $s_2$ if the fitness values of $s_1$  are better, for all the objectives, to the values of $s_2$. In the same way, a solution $s'_1$ non-dominates another solution $s'_2$ if the fitness values of $s'_1$  are better for a subset of objectives and worse to the others. Secondly, the Pareto optimal front is the set of solutions that are non-dominated by other solutions. Therefore, each solution in the Pareto front optimizes one or more objectives, but not all of them, in comparison to the other solutions in the front. Finally, the crowding distance is a metric to estimate the density of solutions surrounding a given one. The crowding distance is calculated obtaining the average distance along each of the objectives of the closest neighbors of the solution.

The NSGA-II algorithm selection operator is based on the idea that the probability of selecting a solution would be higher when it is a non-dominated solution and it has less solutions surrounding it. Therefore, the NSGA-II algorithm sorts the solutions in front levels, and all the solutions in the same front level are ordered using the crowding distance (lines~\ref{alg_front} to~\ref{alg_order} in Algorithm~\ref{algorithm}). Once all the solutions are sorted, it applies a binary tournament selection operator over the sorted elements (lines~\ref{alg_sel1} to~\ref{alg_sel2}): two solutions are selected randomly and the first one in the ordered list is finally selected~\cite{konak2006multi}.

The front levels are generated iteratively. In a first step, the Pareto optimal front is calculated, and it is considered the first front level. In a second iteration, all the solutions in this first front level are removed, and the new Pareto optimal front is calculated, obtaining the second front level. This process is repeated until all the solutions are include in a front level.

\subsection{Offspring generation}

The offspring generation is related on how the individuals of the new population are selected. In the case of NSGA-II, an elitism technique is used~\cite{konak2006multi}, where the new individuals are only included in the new generation if they improve the most dominant solutions.

In each algorithm iteration, a set of new individuals, with the same size than the population, is generated (lines~\ref{alg_beg_for} to~\ref{alg_end_for} in Algorithm~\ref{algorithm}). These new individuals are joined with the individuals in the parent population (line~\ref{alg_join}), obtaining a set of the double size of the population. Only the half of the individuals need to be selected for the offspring. The front levels and crowding distances of the new set are calculated and used to order the solution set (lines~\ref{alg_front} to~\ref{alg_order}). The best half of the solutions is selected as individuals of the offspring (line~\ref{alg_half}), and the rest are rejected.

\subsection{Execution parametrization}

Finally, the parametrization of the algorithm is required. The values of these parameters also influence the performance of the algorithm. Previous testing and/or bibliography research need to be done to establish them.

The parametrization is related to the starting and ending conditions of the algorithm and how often some of the chromosome operators are applied. In our case, the initial population is generated randomly (line~\ref{alg_rand_gen} in Algorithm~\ref{algorithm}). The number of individuals in the population is fixed to 200 solutions. The ending condition is established by the number of generations. We fix this value in 300 generations. Finally, the crossover probability is fixed to 1.0 because the design of NSGA-II already considers the possibility of keeping fathers from one generation to the offspring. The mutation probability is establish in 0.25 and the selection of the mutation is done uniformly. Table~\ref{gaparameters} summarizes the parameters for the execution of the genetic algorithm.

\begin{table}
	\caption{Genetic algorithm execution parameters}
	\label{gaparameters}       % Give a unique label
	\begin{tabular}{lr}
		\hline\noalign{\smallskip}
		Parameter & Value \\
		\noalign{\smallskip}\hline\noalign{\smallskip}
		Population size & 200  \\
		Generations number & 300 \\
		Mutation probability & 0.25 \\
		Mutation selection probability & $1/3$ \\
		Crossover probability & $1.0$ \\
		\noalign{\smallskip}\hline
	\end{tabular}
\end{table}

\begin{algorithm}
	\caption{Multi-objective optimization algorithm}
	\label{algorithm}
	\begin{algorithmic}[1]
		\Procedure{NSGA-II}{}
		\State $populationSize \gets 200$
		\State $generationNumber \gets 300$
		\State $mutationProb \gets 0.25$
		\State $P_t \gets generateRandomPopulation(populationSize)$ \label{alg_rand_gen}
		\State $fitness \gets calculateFitness(P_t)$
		\State $fronts \gets calculateFronts(P_t,fitness)$
		\State $distances \gets calculateCrowding(P_t,fronts,fitness)$
		\For{$i < generationNumber$}
		\State $P_{off} = \emptyset$ 
		\For{$j < populationSize$} \label{alg_beg_for}
		\State $father1 \gets binaryTournamentSelection(P_t,fronts,distances)$ \label{alg_sel1}
		\State $father2 \gets binaryTournamentSelection(P_t,fronts,distances)$ \label{alg_sel2}
		\State $child1,child2 \gets crossover(father1,father2)$
		\If {$random() < mutationProb$} $mutation(child1)$
		\EndIf
		\If {$random() < mutationProb$} $mutation(child2)$
		\EndIf
		\State $P_{off} = P_{off} \cup \{child1, child2\} $ 
		\EndFor \label{alg_end_for}
		\State $P_{off} = P_{off} \cup P_{t}$  \label{alg_join}
		\State $fitness \gets calculateFitness(P_t)$ \label{alg_front}
		\State $fronts \gets calculateFronts(P_{off},fitness)$ 
		\State $distances \gets calculateCrowding(P_{off},fronts,fitness)$
		\State $P_{off} = orderElements(P_{off},fronts,distances)$ \label{alg_order}
		\State $P_{t} = P_{off}[1..populationSize]$ \#the best half \label{alg_half}
		
		\EndFor
		\State $Solution = fronts[1]$ \#the Pareto front
		
		\EndProcedure
	\end{algorithmic}
\end{algorithm}

\section{Experimental evaluation}
In our experiments, we measured the values of the multi-objective functions for different levels of workload and cluster capacity and the number of generations needed to find suitable values. Our results were also compared with another approach.

The first step for the execution of the experiments was to set up the values for the parameters of our model. The access to real data from enterprise's applications is not alway straight forward. Thus, the parameters for the model were estimated from scenarios as similar as possible to our application case. Tables~\ref{experimentsetup} and~\ref{microservicestack} show the parameters values for system and workload models and for the microservices stack, respectively.

The selection of the parameters values for the microservices were based on the analysis of \textit{Socks Shop}~\cite{weaveworks2016shocksshop}. \textit{Socks Shop} is a microservices demo application that was developed to test the benefits of deploying applications in a container platform. The demo also includes a load test that we used to measure the computational resources consumption for each microservice by benchmarking a live demo\footnote{http://cloud.weave.works/demo}. The values for the failure rates of the microservices were estimated using data from cloud application traces~\cite{di2015gloudsim} and the size of the source code of \textit{Socks Shop}\footnote{https://github.com/microservices-demo}\footnote{https://hub.docker.com/u/weaveworksdemos/}. The number of microservices requests for each application request was estimated from the customer behavior model graph (CBMG) of an e-commerce benchmark~\cite{menasce2002tpc}.

We considered a heterogeneous cluster with four types of physical machines that differed between them in their computational capacity: $cap=[100.0,$ $ 200.0,$ $ 400.0,$ $ 800.0]$. The four machines types were fixed with the same value for the fail rate, $fail=0.025$~\cite{fu2010failure}. The experiments contained the same quantity of machines of each type. The physical machines were distributed in two racks. The network distance for physical machines in the same rack was fixed in $d=1.0$. Network distance for machines in different racks was fixed in $d=4.0$. 

In order to study our approach under different conditions, we executed the optimization process for several experiment configurations. The workload of the system was varied between the experiments by modifying the number of applications running in the system ($|A|=[1, 2]$) and by changing the number of user requests that an application received ($ureq_j=[1.0, 1.5, 2.0]$). The capacity of the cluster was modified by changing the number of available physical machines ($|P|=[250, 300, 350, 400]$). Thus, our experimentation included 24 configurations.

\begin{table}
	\caption{Experiments set up}
	\label{experimentsetup}       % Give a unique label
	\begin{tabular}{lll}
		\hline\noalign{\smallskip}
		Element & Parameter & Value  \\
		\noalign{\smallskip}\hline\noalign{\smallskip}
		Application   & $|A|$ & 1 and 2 \\
		& $ureq_j$ & 1.0, 1.5 and 2.0 \\
		Physical machine      	& $|P|$ & 250, 300, 350 and 400\\
		& $cap_l$ & 100.0, 200.0, 400.0 or 800.0\\
		& $fail_l$ & 0.025\\
		Network       & $dist_{pm_{l},pm_{l'}}$ & 1.0 or 4.0\\
		Microservice& &  Table~\ref{microservicestack} \\
		
		\noalign{\smallskip}\hline
	\end{tabular}
\end{table}

\begin{table}
	\caption{Microservices stack for \textit{Socks Shop} application}
	\label{microservicestack}       % Give a unique label
	\begin{tabular}{lrrrrrr}
		\hline\noalign{\smallskip}
		Name & Id. & Consumes & $msreq_i$ & $res_i$ & $thr_i$ & $fail_i$  \\
		\noalign{\smallskip}\hline\noalign{\smallskip}
		worker & $ms_{1}$ & $\{\}$ & 3.2 & 0.1 & 1.0 & 0.04 \\		
		shipping & $ms_{2}$ & $\{ms_{13}\}$ & 1.8 & 11.7 & 25.0 & 0.02 \\		
		queue-master & $ms_{3}$ & $\{s_2,s_4\}$ & 3.2 & 20.0 & 200.0 & 0.02 \\		
		payment & $ms_{4}$ & $\{\}$ & 1.4 & 0.1 & 10.0 & 0.0002\\		
		orders & $ms_{5}$ & $\{ms_{2},ms_{4},ms_{10},ms_{11},ms_{12},\}$ & 2.3 & 27.1 & 80.0 & 0.02 \\		
		login & $ms_{6}$ & $\{\}$ & 0.8 & 2.8 & 30.0 & 0.0001 \\		
		front-end & $ms_{7}$ & $\{ms_{5},ms_{6},ms_{9}\}$ & 15.1 & 3.8 & 50.0 & 0.003 \\		
		edge-router & $ms_{8}$ & $\{ms_{7}\}$ & 15.1 & 0.5 & 10.0 & 0.0001 \\		
		catalogue & $ms_{9}$ & $\{\}$ & 12.0 & 0.2 & 3.0 & 0.0006 \\		
		cart & $ms_{10}$ & $\{ms_{12}\}$ & 3.2 & 41.3 & 100.0 & 0.02 \\		
		accounts & $ms_{11}$ & $\{ms_{12}\}$ & 0.1 & 45.1 & 100.0 & 0.003 \\		
		weavedb & $ms_{12}$ & $\{\}$ & 3.2 & 26.3 & 80.0 & 0.04 \\		
		rabbitmq & $ms_{13}$ & $\{ms_{1},ms_{3}\}$ & 3.2 & 4.0 & 40.0 & 0.0006 \\		
		consul & $ms_{14}$ & $\{\}$ & 3.2 & 13.2 & 100.0 & 0.0003 \\		
		\noalign{\smallskip}\hline
	\end{tabular}
\end{table}

We did not considered the similar researches explained in Section~\ref{} to compare our solution with, because their differences against our approach. Therefore, the comparison of our results was done with Kubernetes container policies.

Kubernetes uses two policies to manage the allocation of containers or pods, as they are called in Kubernetes implementation: PodFitResources and LeastRequestPriority~\cite{vohra2017scheduling}. The first one limits that the sum of requested resources on a physical machine must not be greater than the capacity of the machine. The second one determines that containers are allocated in machines with least consumed resources. Both policies are focused on spreading the containers across all the machines in the cluster. These policies can not manage the scale level of the containers. In order to a fairer comparison of the results, the scale level of the experiments for Kubernetes was fixed at the same values that our approach obtained in the experiment results. The other experiment configurations and conditions were exactly the same than in the experiments for our approach. 

\section{Results}

The results were analyzed from two perspectives: the adaptation of the solutions to the objectives functions as new generations were obtained (evolution of the objectives functions values) and the results of the optimization ---final values for the optimization objectives---. In the case of the first perspective, a summary of three representative cases is presented.

It is important to remember that the solution obtained by a multi-objective optimization evolutionary algorithm in general, and the NSGA-II in particular, is not a single solution but a set of solutions that correspond to all the individuals placed in the Pareto optimal front. In favor of simplifying the analysis of the results, three data series are represented: (a) the minimum value of an objective between all the solutions in the Pareto front, labeled as \textit{min}; (b) the mean of all the values of an objective for all the solutions in the Pareto front, labeled as \textit{mean}; and (c) the value of the objective for a representative solution in the Pareto front, labeled as \textit{minSOV}. The selection of the representative solution was done by choosing the one with the smaller single objective value (SOV). This SOV was calculated as the weighted average of the normalized values of the four objectives (Equation~\ref{eq:sov}). Our criteria for selecting a solution was arbitrary and it was done only with the propose of giving more details in the analysis of the results. The criteria to choose between all the set of solutions with the same quality, the Pareto front, will depend on the conditions and requirements of the specific environment where the solution will be implemented. For instance, an environment in which reliability of the system is more important than other objectives, solutions in the Pareto front with smaller values for the System Failure objective should be selected.

\begin{eqnarray}
\label{eq:sov}
SOV = 0.25 \frac{value_{threshold}-min_{threshold}}{max_{threshold}-min_{threshold}} + 0.25 \frac{value_{cluster}-min_{cluster}}{max_{cluster}-min_{cluster}} \\
+ 0.25 \frac{value_{failure}-min_{failure}}{max_{failure}-min_{failure}} + 0.25  \frac{value_{network}-min_{network}}{max_{network}-min_{network}}
\end{eqnarray}

\begin{figure*}
	\includegraphics[width=0.48\textwidth]{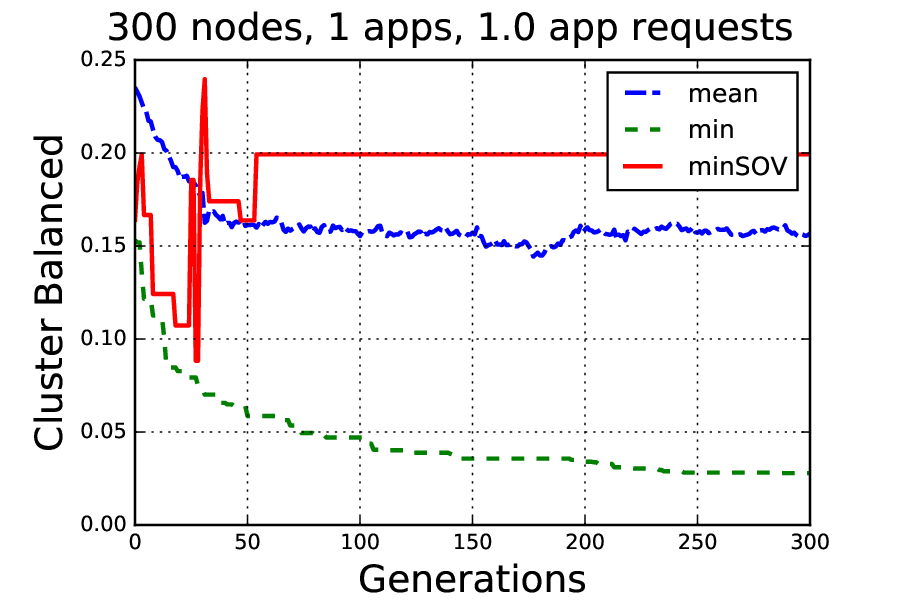}
	\includegraphics[width=0.48\textwidth]{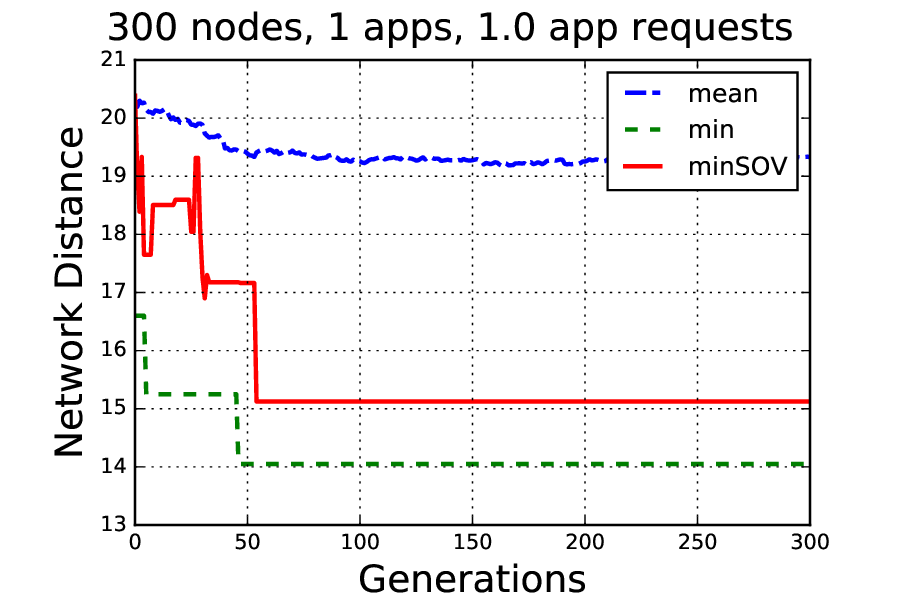}\\
	\includegraphics[width=0.48\textwidth]{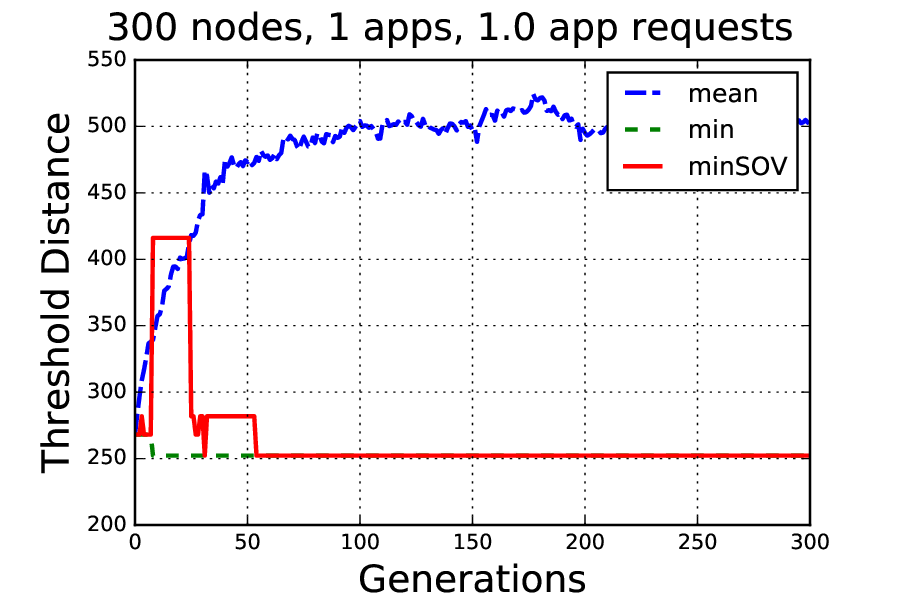}
	\includegraphics[width=0.48\textwidth]{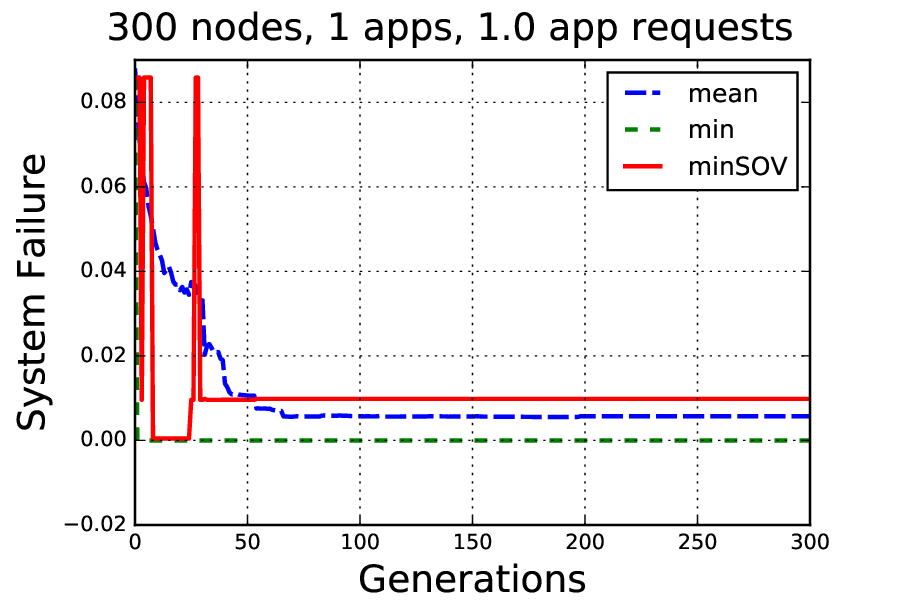}\\	
	\caption{Evolution of the four objectives functions values along the generations for the experiment with 1 application, with a workload level of 1.0 user requests using 300 physical machines}
	\label{fig:objvalues300-10-1}       % Give a unique label
\end{figure*}

\begin{figure*}
	\includegraphics[width=0.48\textwidth]{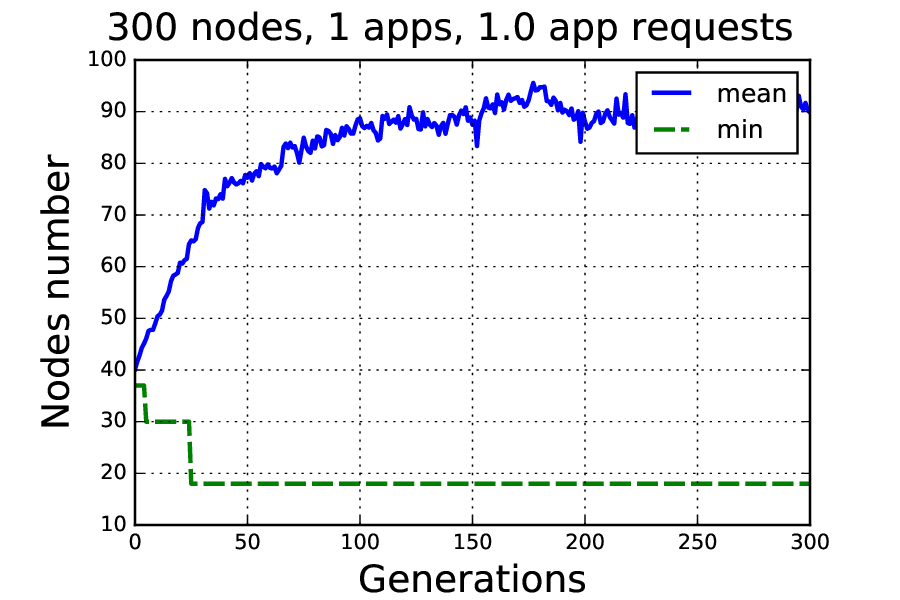}
	\includegraphics[width=0.48\textwidth]{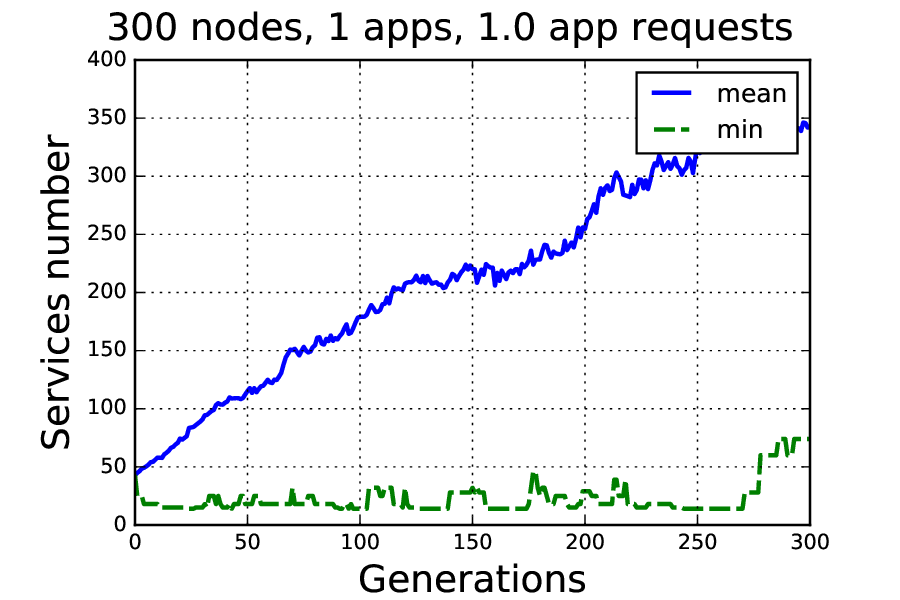}	
	\caption{Evolution of the system state (nodes and services)  along the generations for the experiment with 1 application, with a workload level of 1.0 user requests using 300 physical machines}
	\label{fig:sysvalues300-10-1}       % Give a unique label
\end{figure*}

\begin{figure*}
	\includegraphics[width=0.48\textwidth]{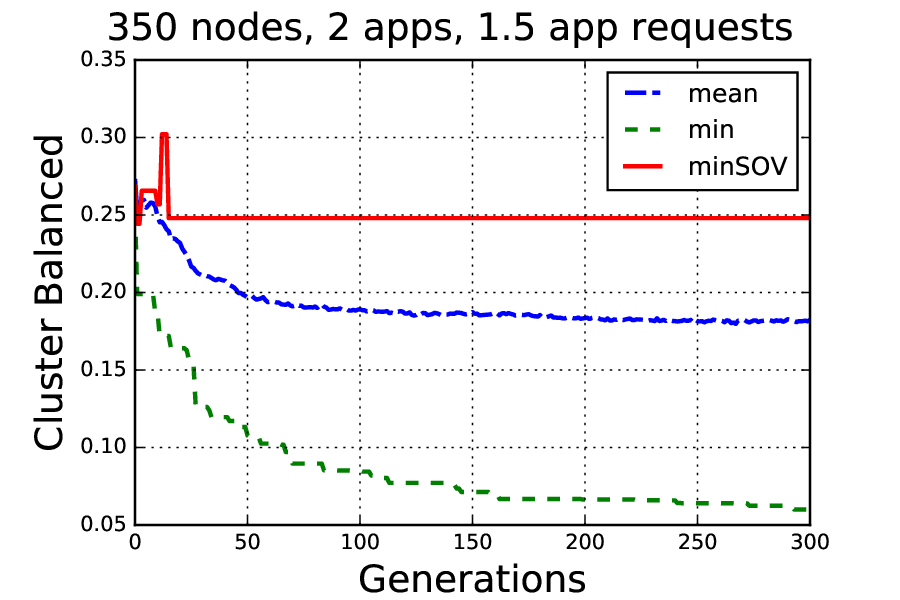}
	\includegraphics[width=0.48\textwidth]{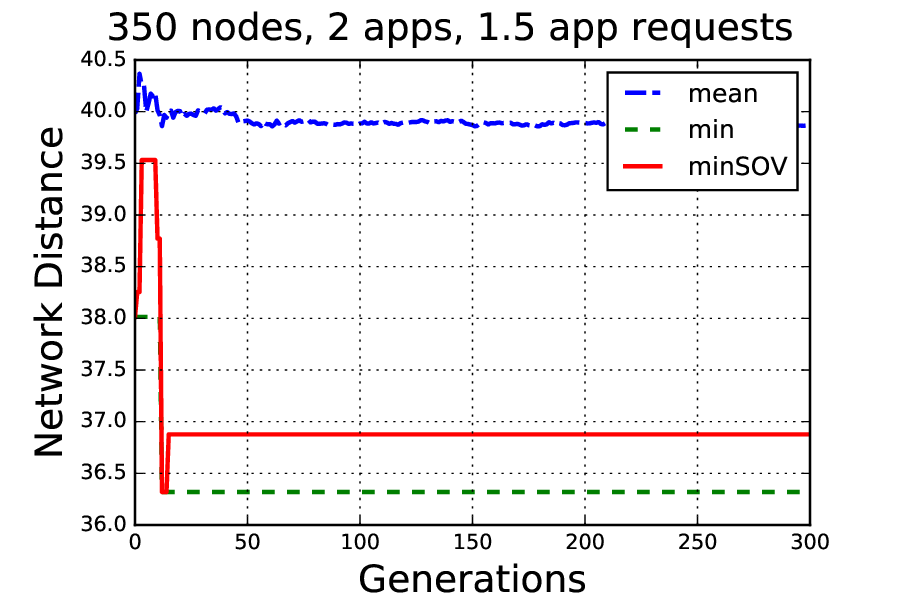}\\
	\includegraphics[width=0.48\textwidth]{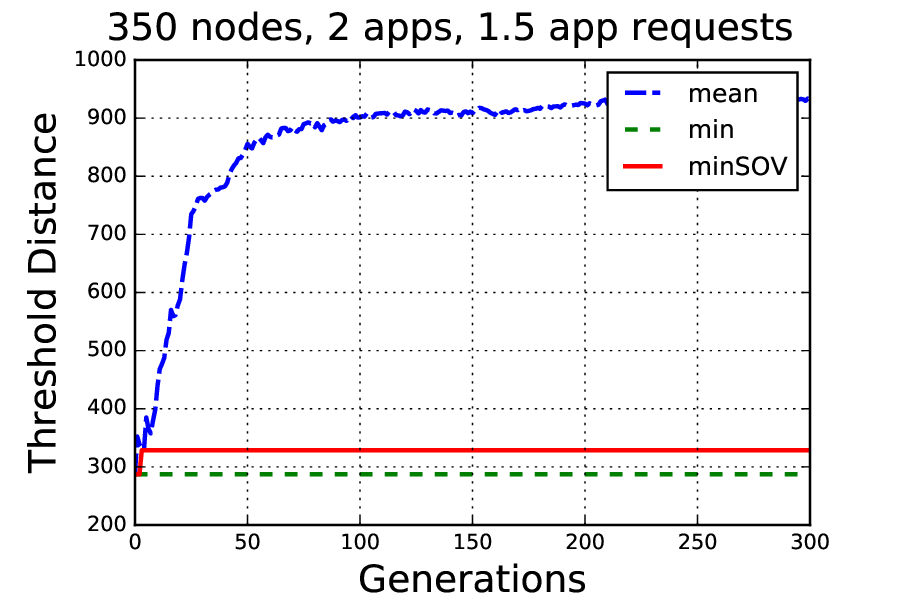}
	\includegraphics[width=0.48\textwidth]{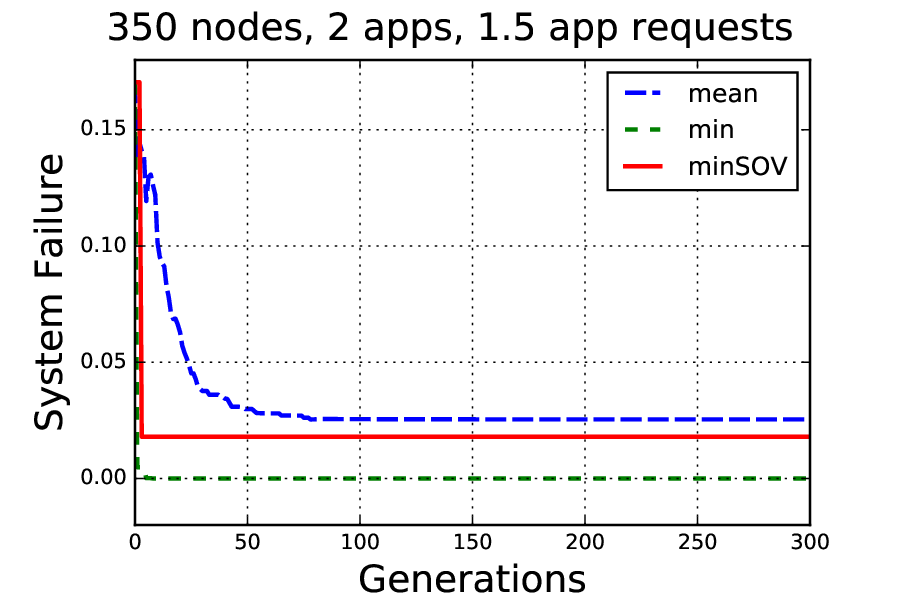}\\
	\caption{Evolution of the four objectives functions values along the generations for the experiment with 2 application, with a workload level of 1.5 user requests using 350 physical machines}
	\label{fig:objvalues350-15-2}       % Give a unique label
\end{figure*}

\begin{figure*}
	\includegraphics[width=0.48\textwidth]{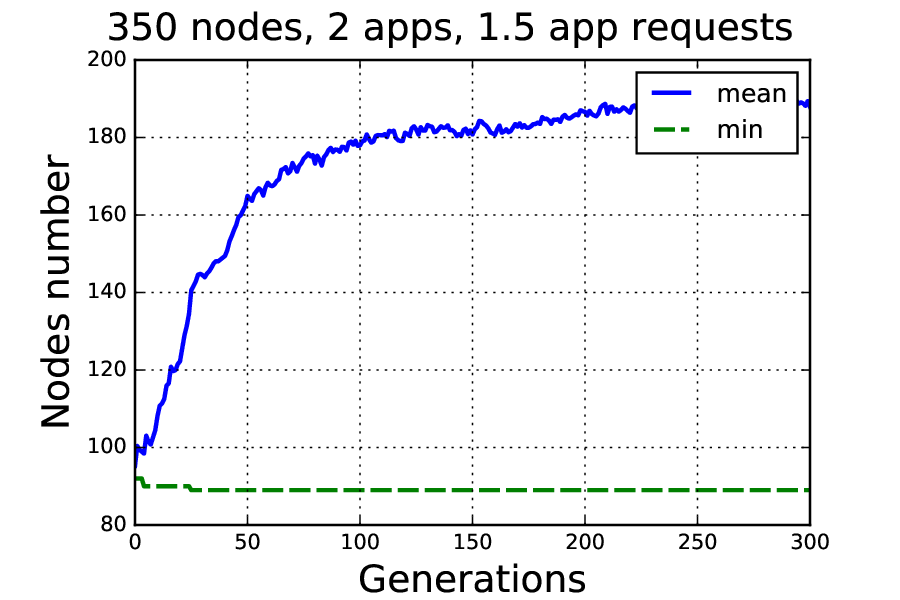}
	\includegraphics[width=0.48\textwidth]{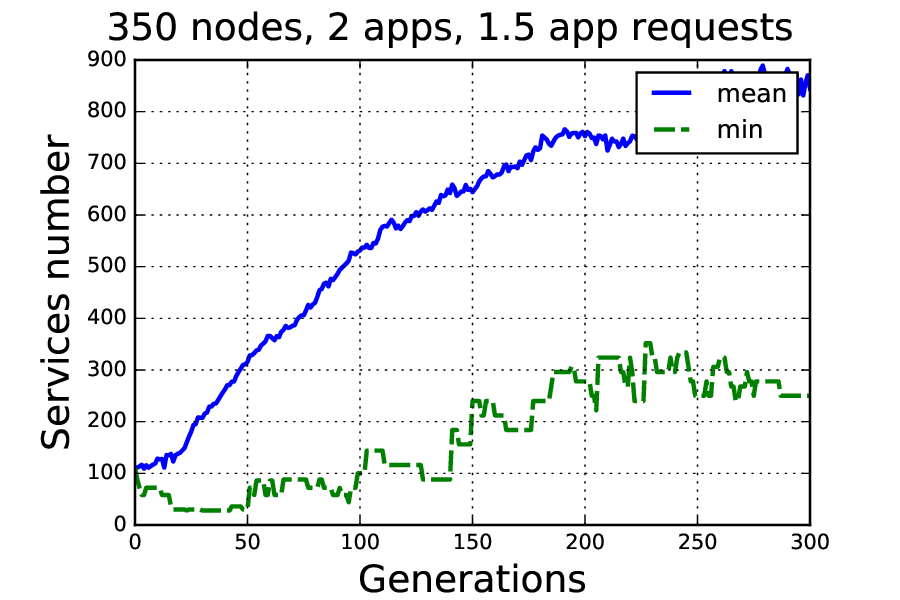}	
	\caption{Evolution of the system state (nodes and services)  along the generations for the experiment with 2 application, with a workload level of 1.5 user requests using 350 physical machines}
	\label{fig:sysvalues350-15-2}       % Give a unique label
\end{figure*}

\begin{figure*}
	\includegraphics[width=0.48\textwidth]{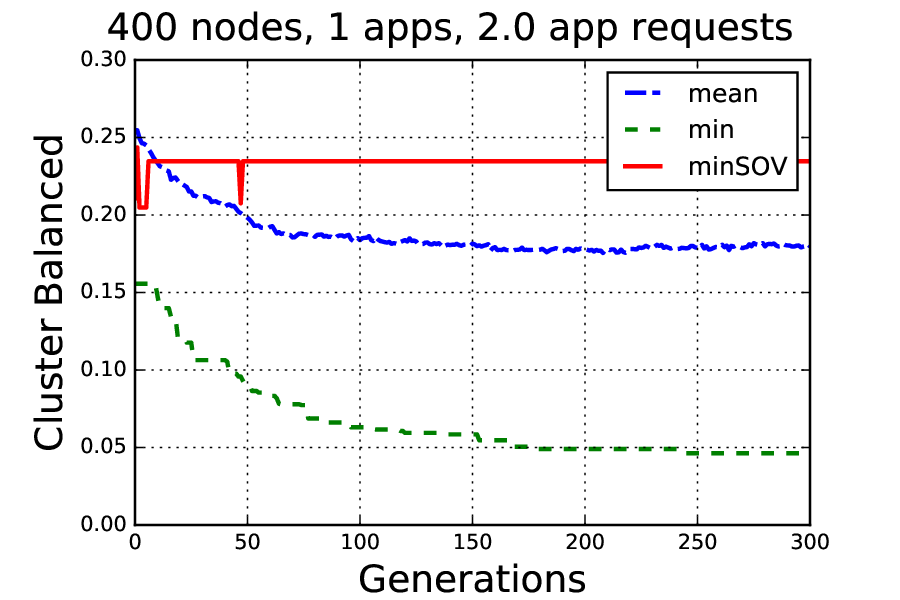}
	\includegraphics[width=0.48\textwidth]{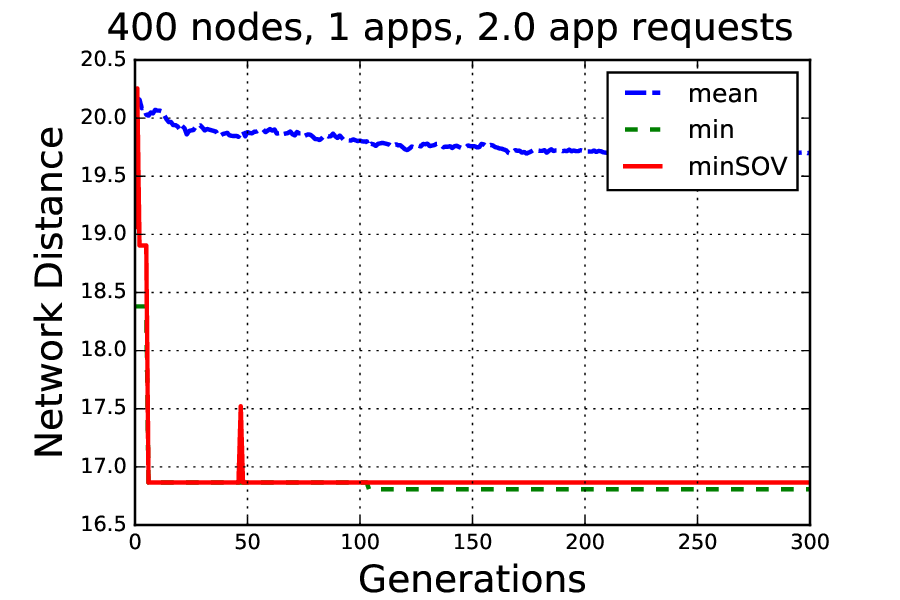}\\
	\includegraphics[width=0.48\textwidth]{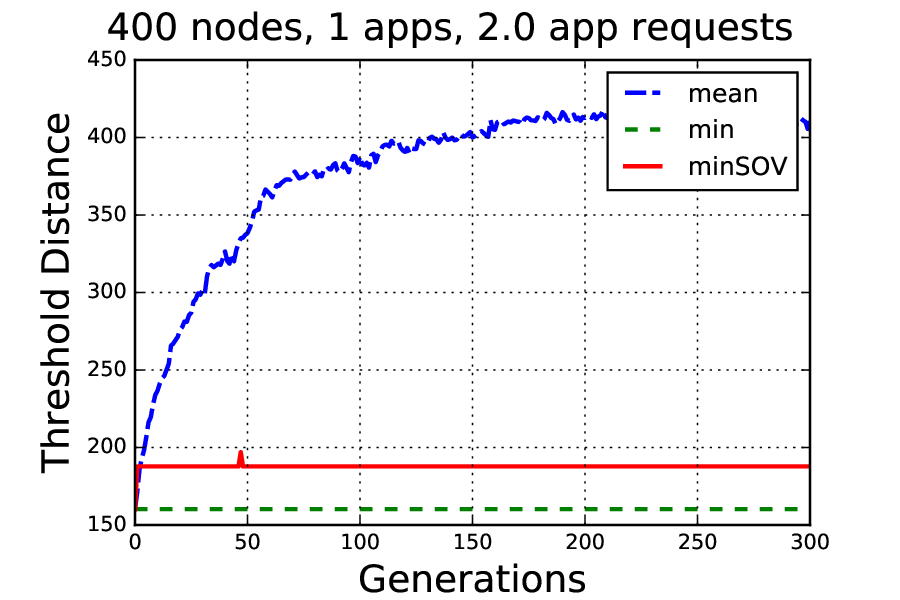}
	\includegraphics[width=0.48\textwidth]{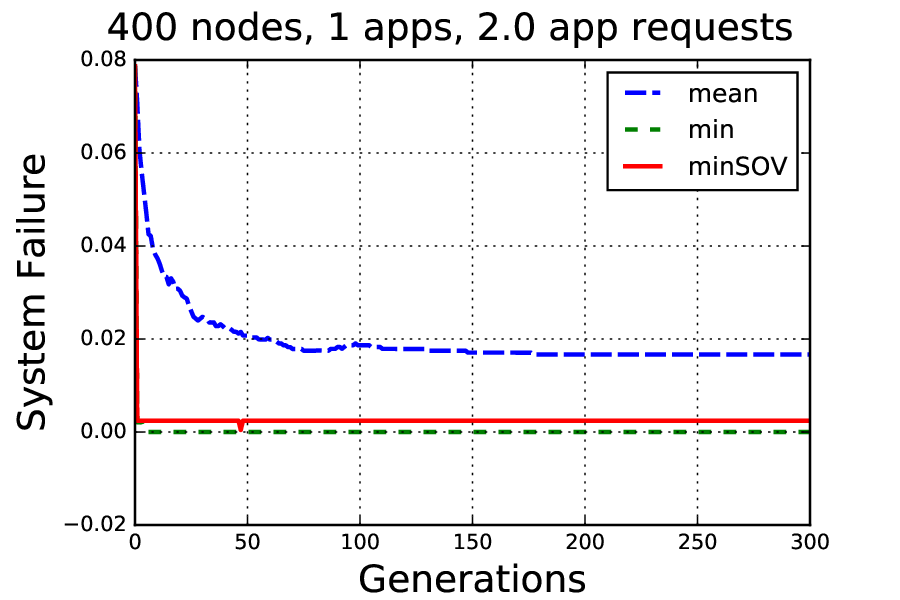}\\
	\caption{Evolution of the four objectives functions values along the generations for the experiment with 1 application, with a workload level of 2.0 user requests using 400 physical machines}
	\label{fig:objvalues400-20-1}       % Give a unique label
\end{figure*}

\begin{figure*}
	\includegraphics[width=0.48\textwidth]{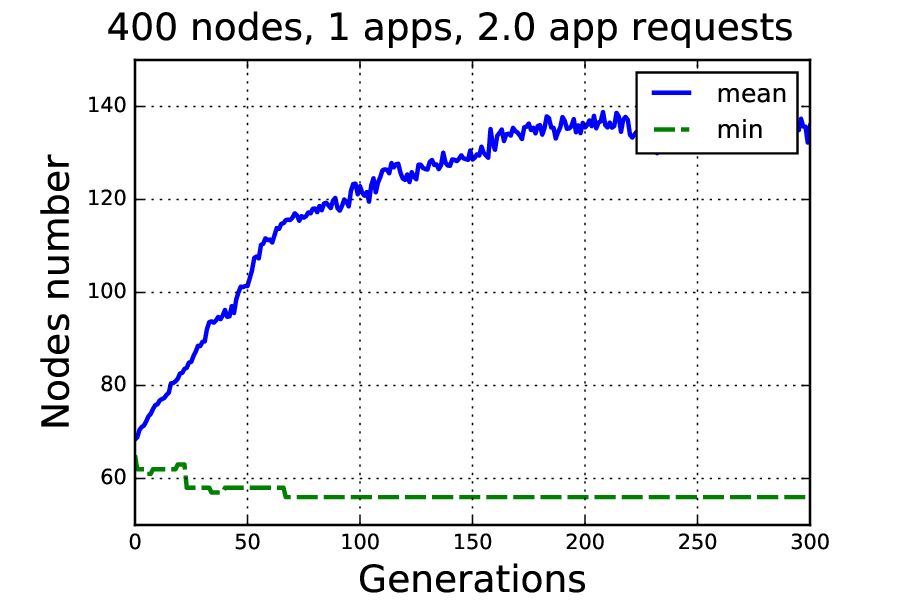}
	\includegraphics[width=0.48\textwidth]{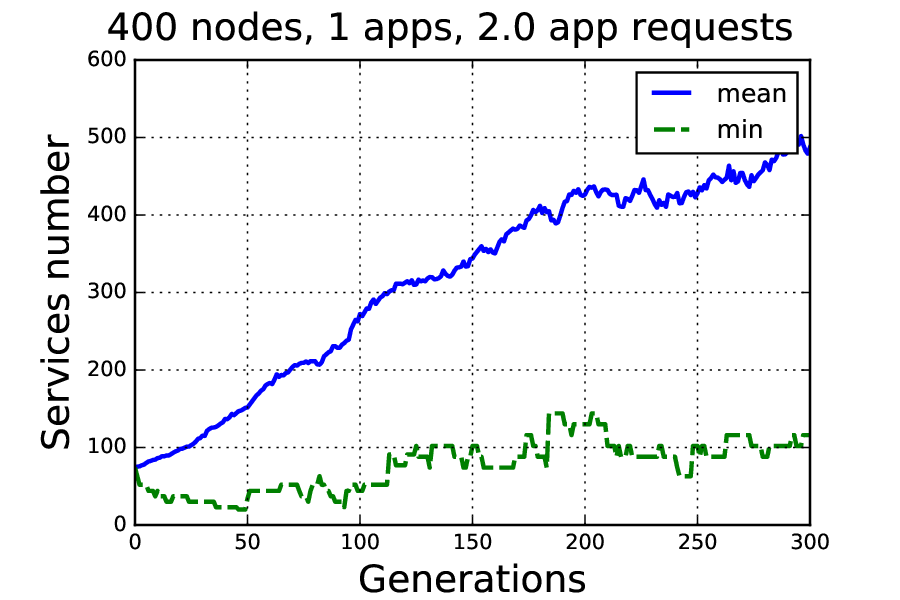}		
	\caption{Evolution of the system state (nodes and services)  along the generations for the experiment with 1 application, with a workload level of 2.0 user requests using 400 physical machines}
	\label{fig:sysvalues400-20-1}       % Give a unique label
\end{figure*}

In the first case, the analysis of the evolution of the solutions, the results for three experiment configurations are presented: Figures~\ref{fig:objvalues300-10-1} and~\ref{fig:sysvalues300-10-1}  for the experiment with 300 physical machines, 1.0 user requests and 1 application; Figures~\ref{fig:objvalues350-15-2} and~\ref{fig:sysvalues350-15-2} for 350 physical machines, 1.5 user requests and 2 applications; and Figures~\ref{fig:objvalues400-20-1} and~\ref{fig:sysvalues400-20-1} for 400 physical machines, 2.0 user requests and 1 application. 

Results are presented separately for the objectives functions ---Figures~\ref{fig:objvalues300-10-1},  \ref{fig:objvalues350-15-2} and~\ref{fig:objvalues400-20-1}, including  the series \textit{min}, \textit{mean} and \textit{minSOV}--- and for the system conditions ---Figures~\ref{fig:sysvalues300-10-1},  \ref{fig:sysvalues350-15-2} and~\ref{fig:sysvalues400-20-1}, including the series \textit{min} and \textit{mean}---. System conditions are represented with microservices scale level and number of physical machines that allocate at least one microservice. 

\begin{figure*}
	\includegraphics[width=0.32\textwidth]{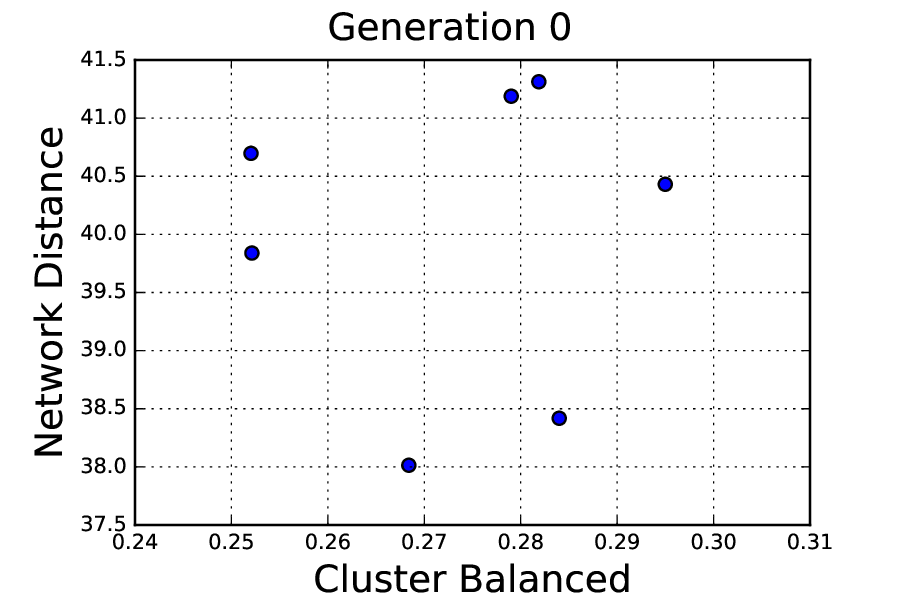}
	\includegraphics[width=0.32\textwidth]{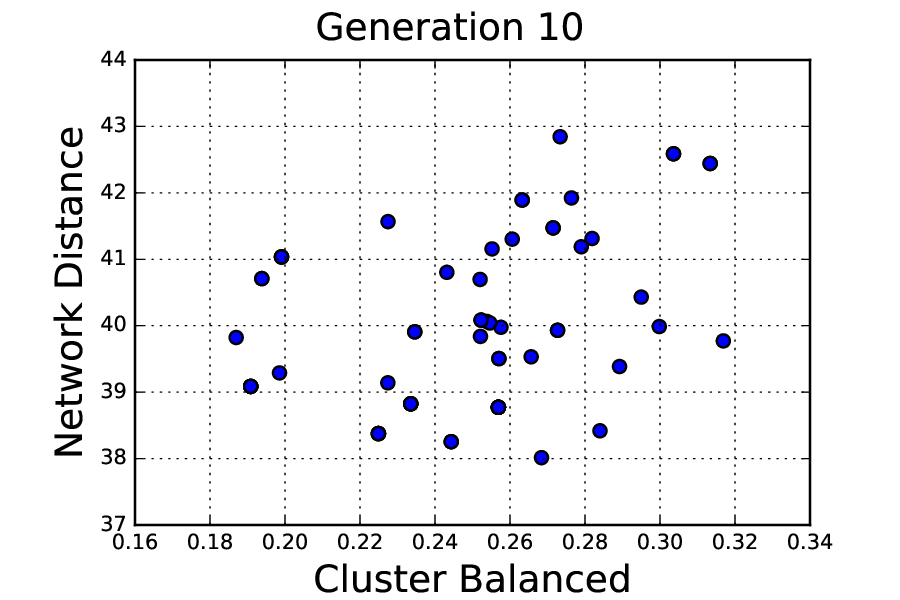}
	\includegraphics[width=0.32\textwidth]{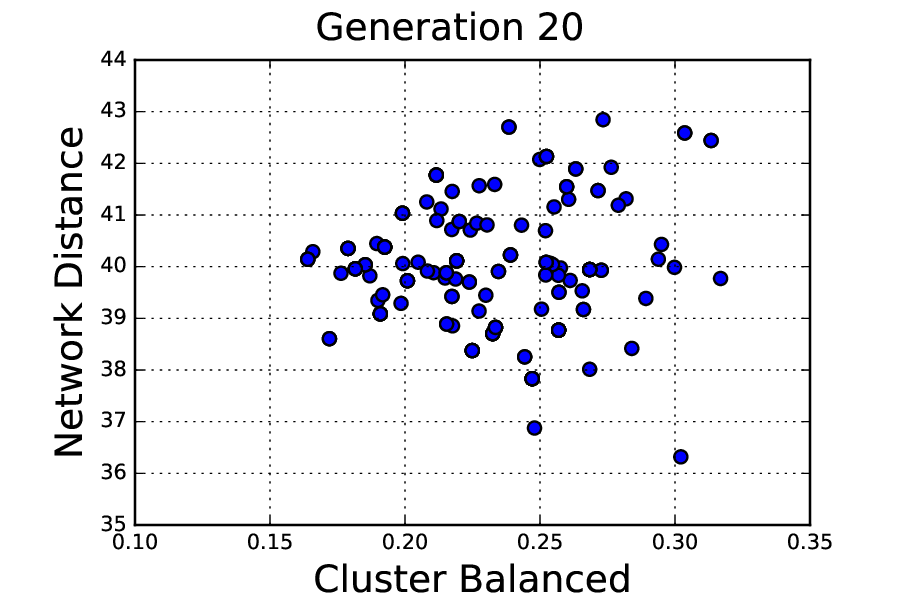}\\
	\includegraphics[width=0.32\textwidth]{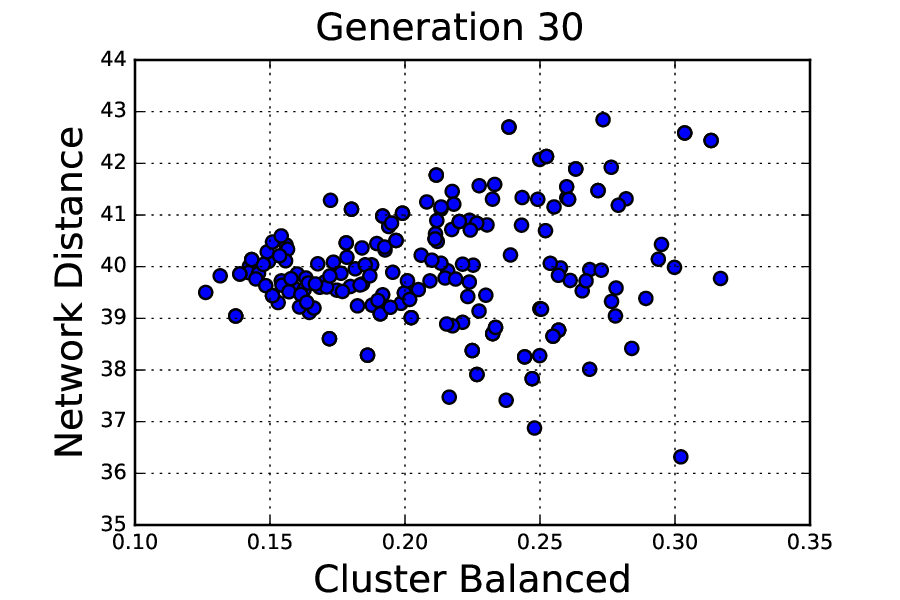}
	\includegraphics[width=0.32\textwidth]{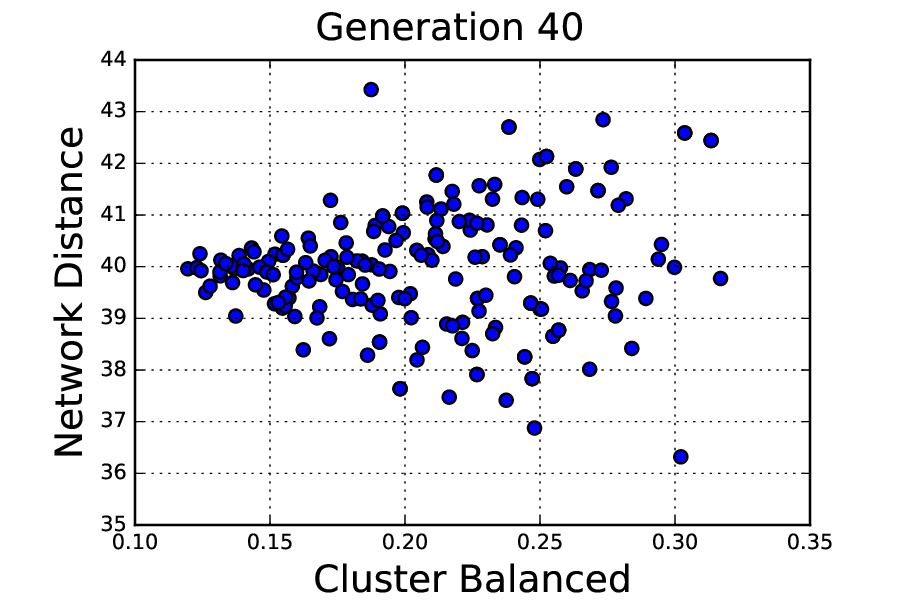}
	\includegraphics[width=0.32\textwidth]{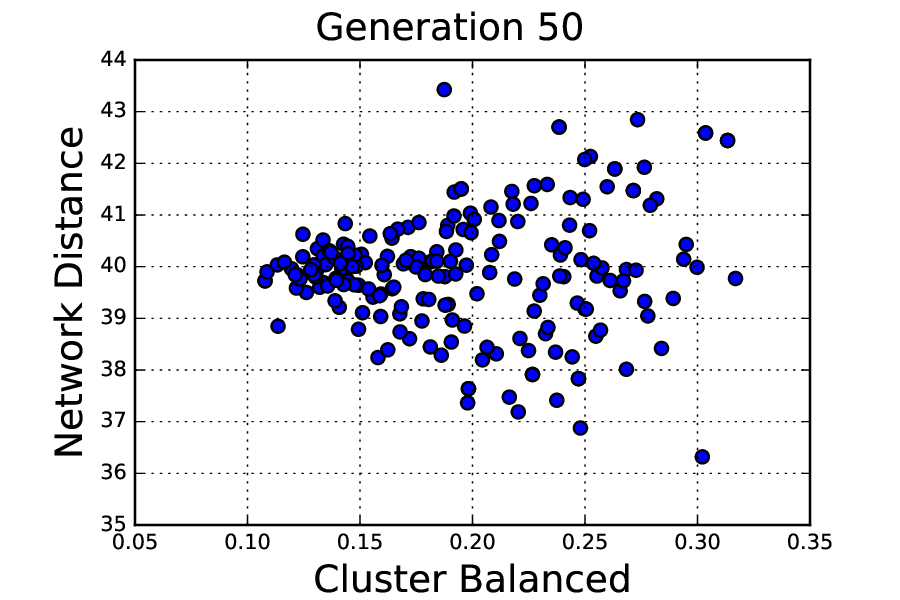}	
	
	\caption{Evolution of the Pareto Optimal Front  along the generations for objectives Network Distance and Cluster Balanced Used for the experiment with 2 applications, 1.5 user requests and 350 physical machines}
	\label{fig:scatternetclus}       % Give a unique label
\end{figure*}

\begin{figure*}
	\includegraphics[width=0.32\textwidth]{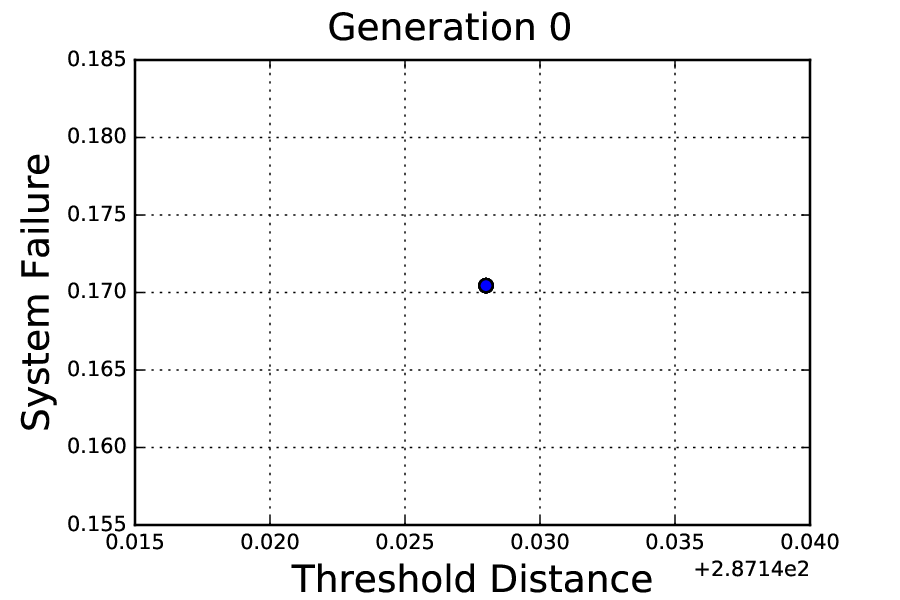}
	\includegraphics[width=0.32\textwidth]{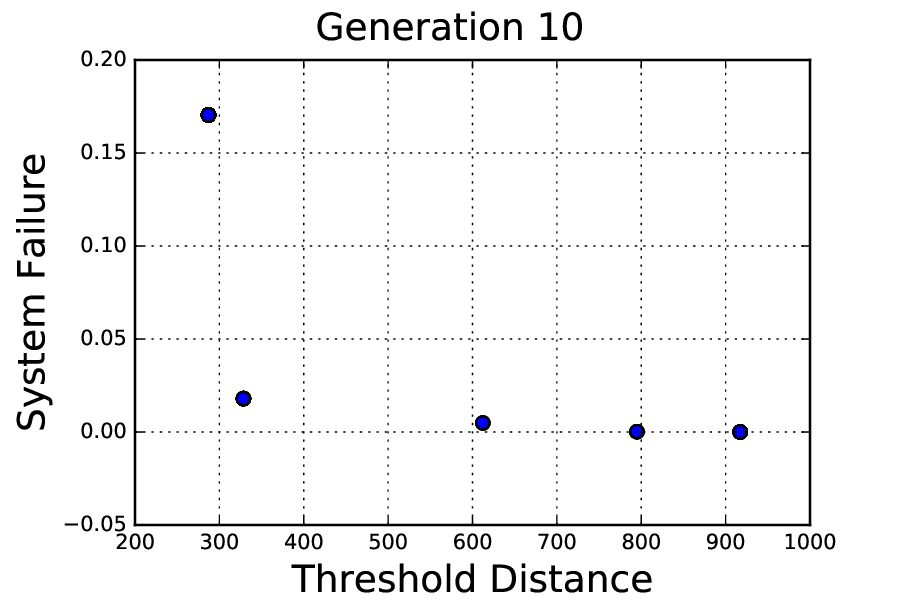}
	\includegraphics[width=0.32\textwidth]{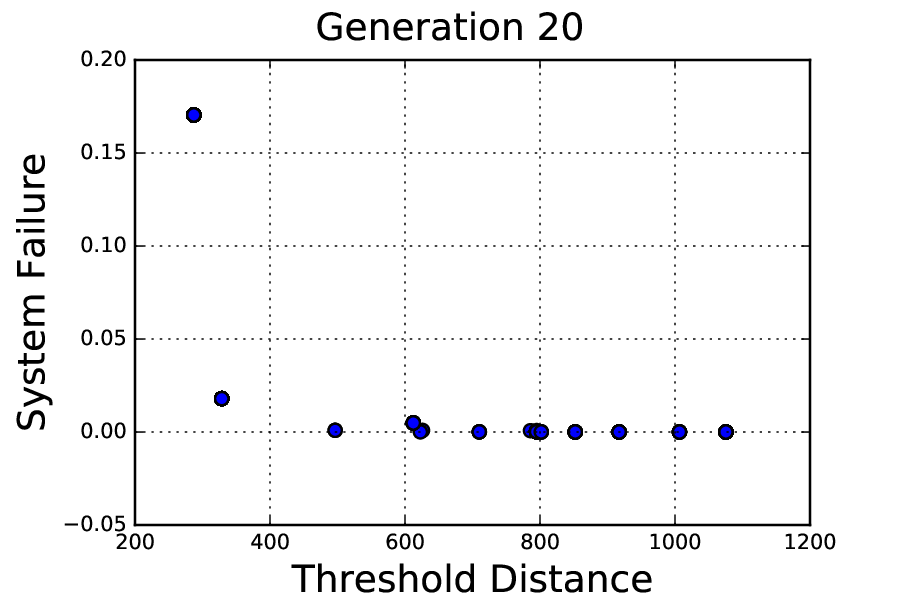}\\
	\includegraphics[width=0.32\textwidth]{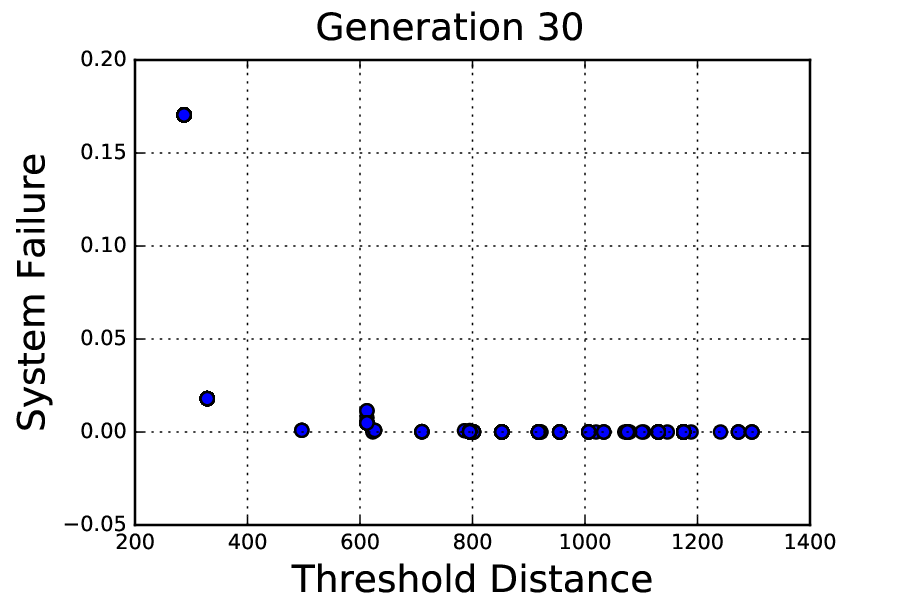}
	\includegraphics[width=0.32\textwidth]{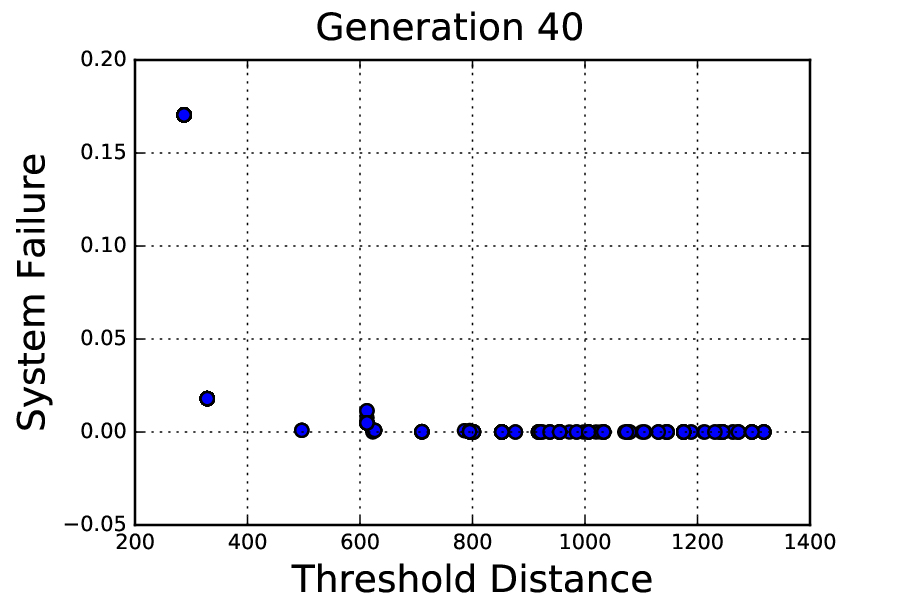}
	\includegraphics[width=0.32\textwidth]{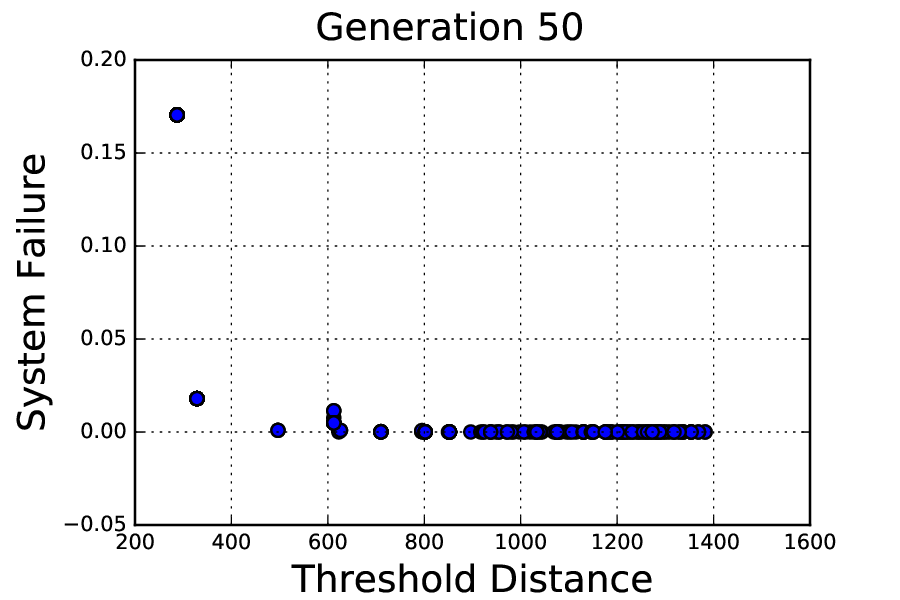}	
	\caption{Evolution of the Pareto Optimal Front  along the generations for objectives System Failure and Threshold Distance for the experiment with 2 applications, 1.5 user requests and 350 physical machines}
	\label{fig:scatterfailthr}       % Give a unique label
\end{figure*}

Finally, one of the previous configurations (350 physical machines, 1.5 user requests and 2 application) is used to illustrate the evolution of all the solutions in the Pareto set (Figures~\ref{fig:scatternetclus} and~\ref{fig:scatterfailthr}). We represent the scatter plots of the values of the objectives functions for six generations. As each of the points in the scatter plot is a four values tuple, one for each objective function, two scatter plots are used for the representation: one for the values of Total Network Distance and Cluster Balanced Used (Figure~\ref{fig:scatternetclus}) and another one for System Failure and Threshold Distance (Figure~\ref{fig:scatterfailthr}). 

\begin{figure*}
	\includegraphics[width=0.48\textwidth]{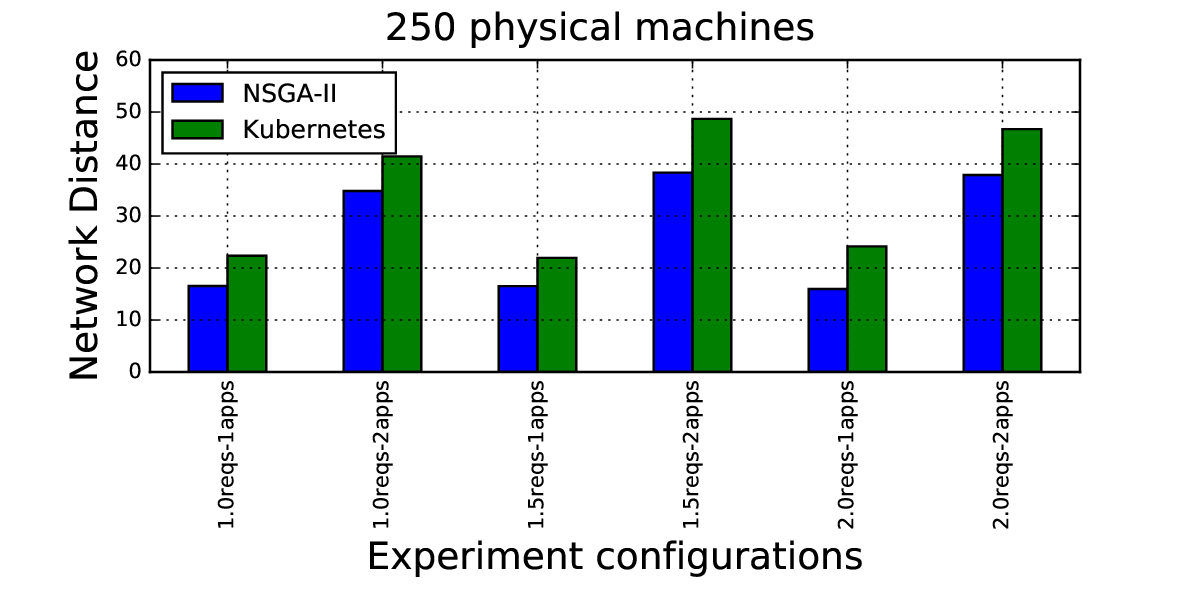}
	\includegraphics[width=0.48\textwidth]{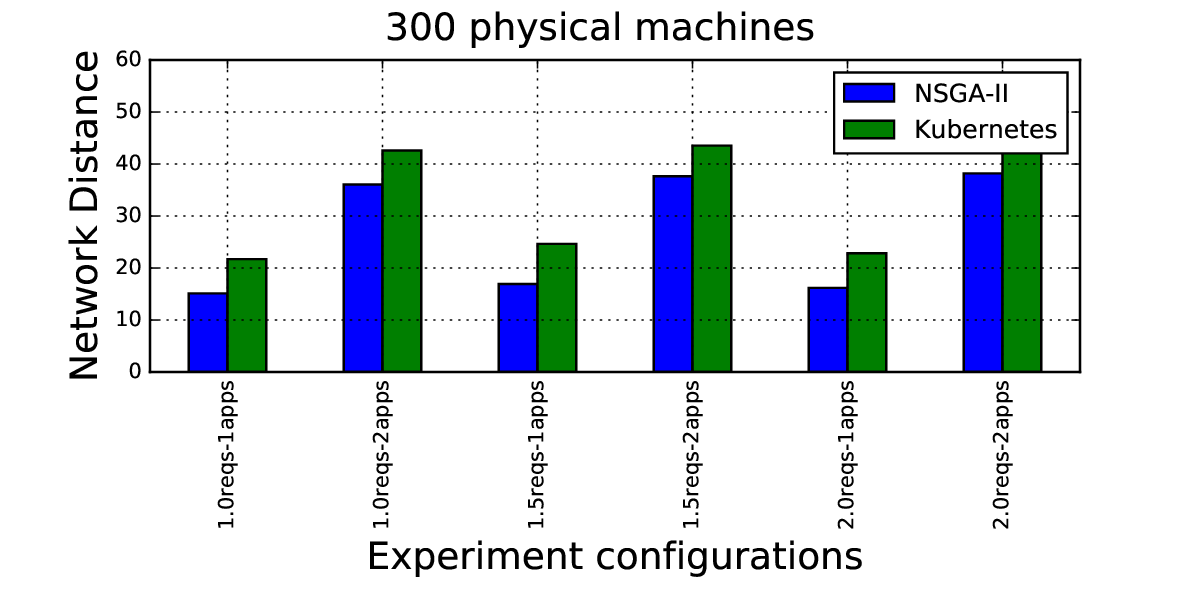}\\
	\includegraphics[width=0.48\textwidth]{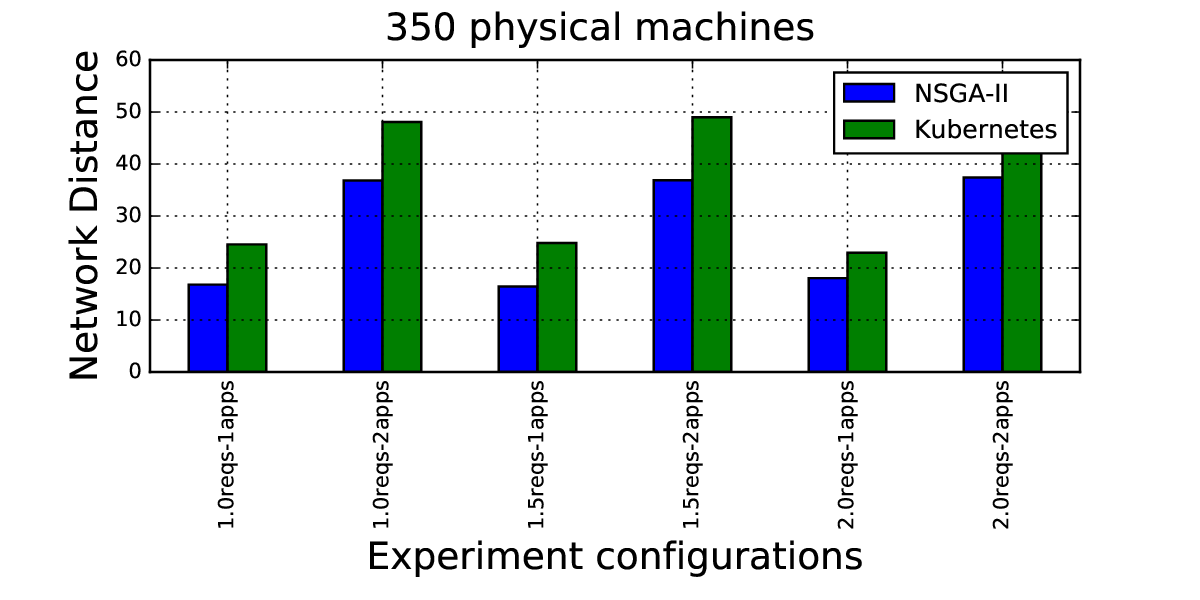}
	\includegraphics[width=0.48\textwidth]{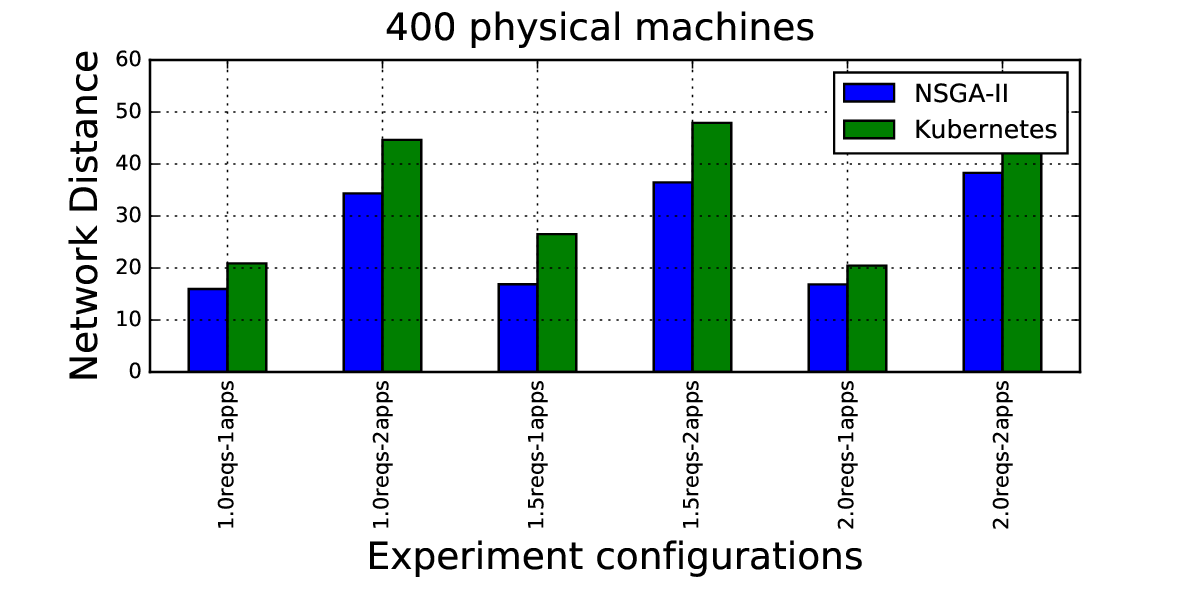}
	\caption{Comparative results for Network Distance optimization}
	\label{fig:comparativeresultsnet}       % Give a unique label
\end{figure*}

\begin{figure*}
	\includegraphics[width=0.48\textwidth]{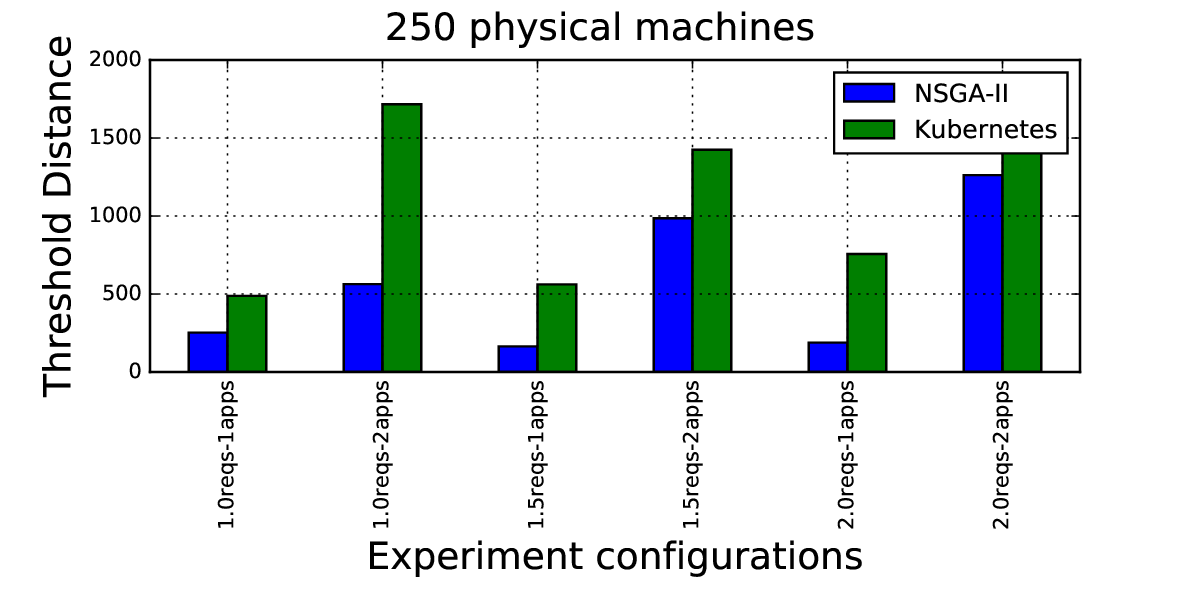}
	\includegraphics[width=0.48\textwidth]{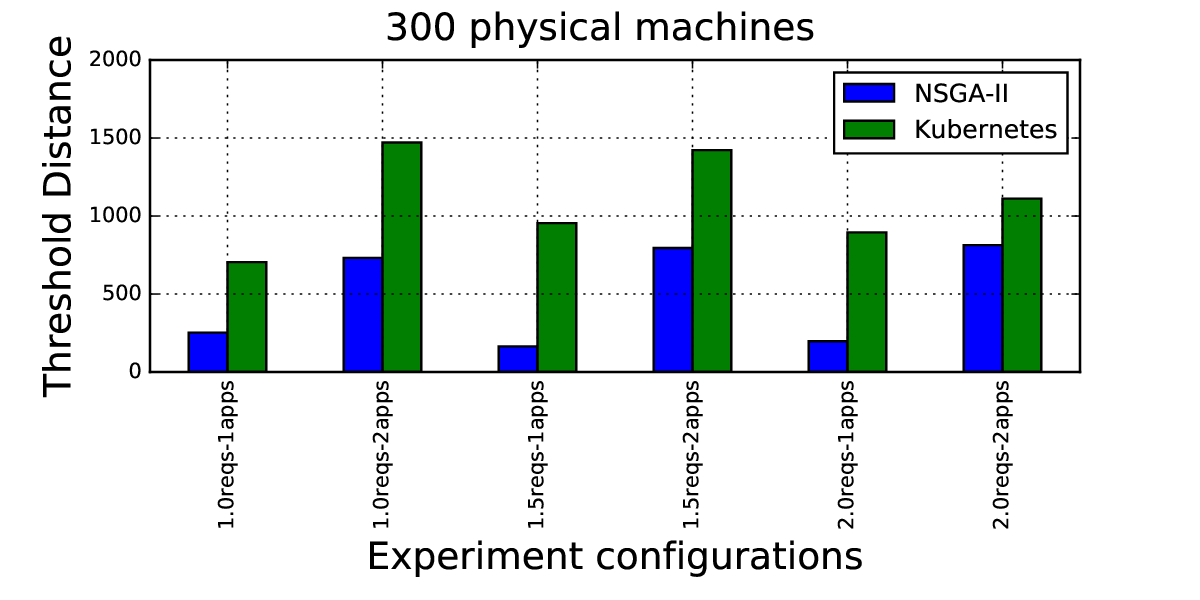}\\
	\includegraphics[width=0.48\textwidth]{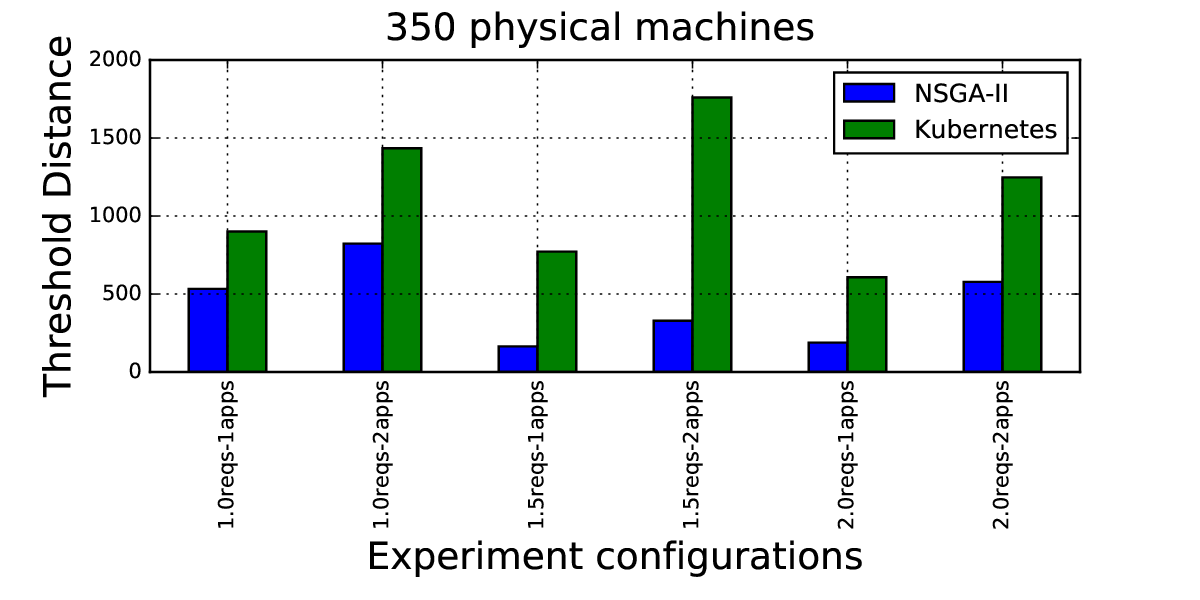}
	\includegraphics[width=0.48\textwidth]{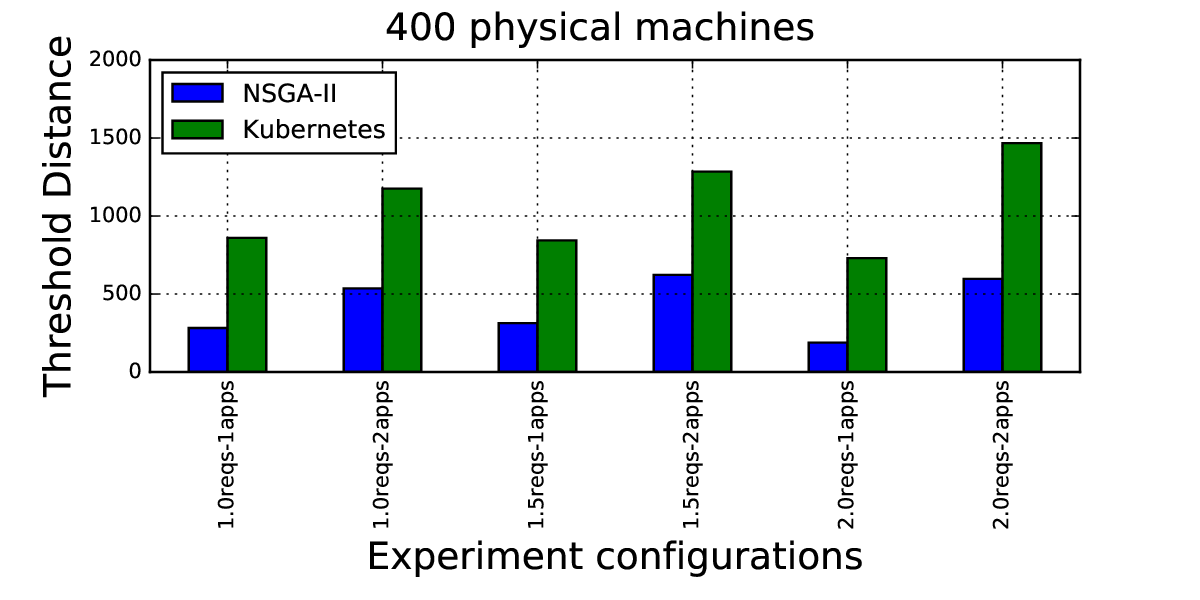}
	\caption{Comparative results for Network Distance optimization}
	\label{fig:comparativeresultsthr}       % Give a unique label
\end{figure*}

\begin{figure*}
	\includegraphics[width=0.48\textwidth]{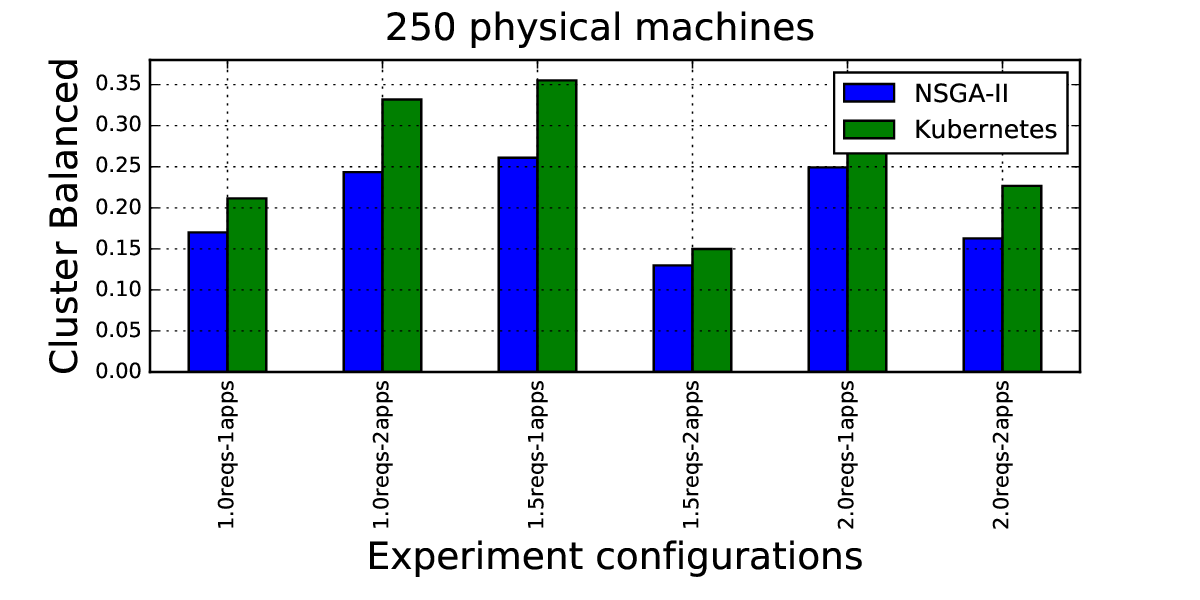}
	\includegraphics[width=0.48\textwidth]{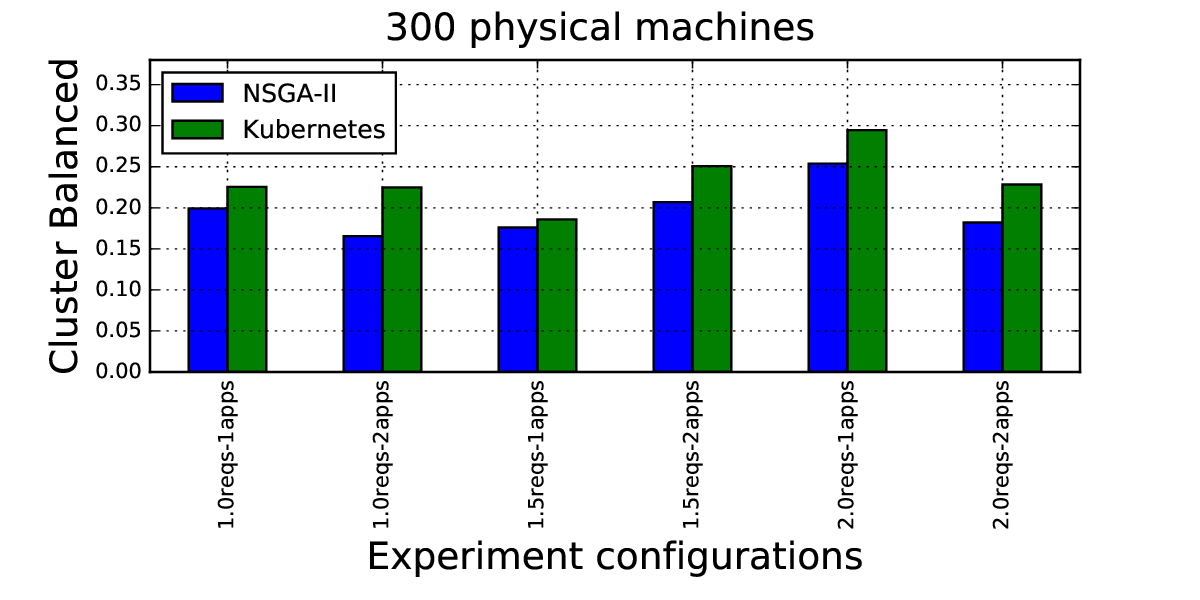}\\
	\includegraphics[width=0.48\textwidth]{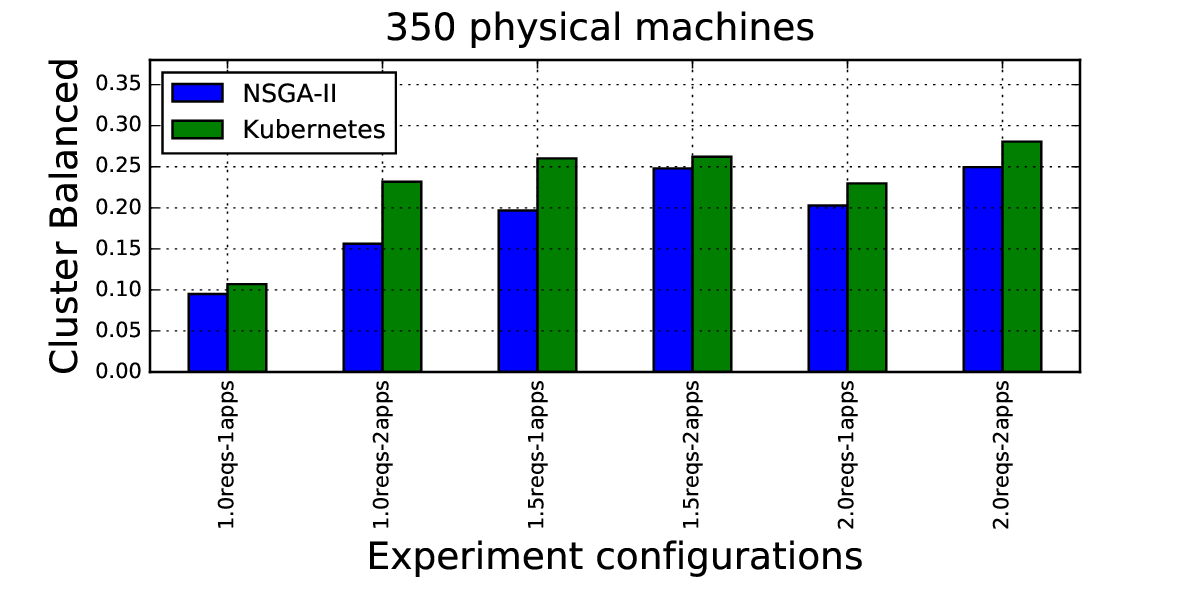}
	\includegraphics[width=0.48\textwidth]{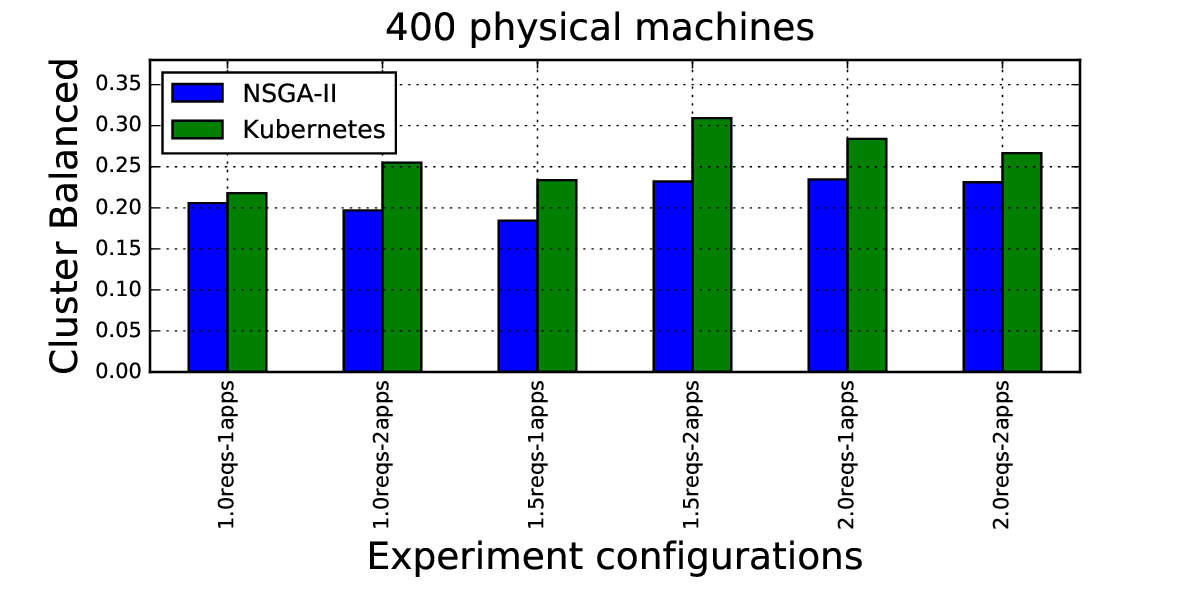}
	\caption{Comparative results for Cluster Balanced Used optimization}
	\label{fig:comparativeresultsclus}       % Give a unique label
\end{figure*}

\begin{figure*}
	\includegraphics[width=0.48\textwidth]{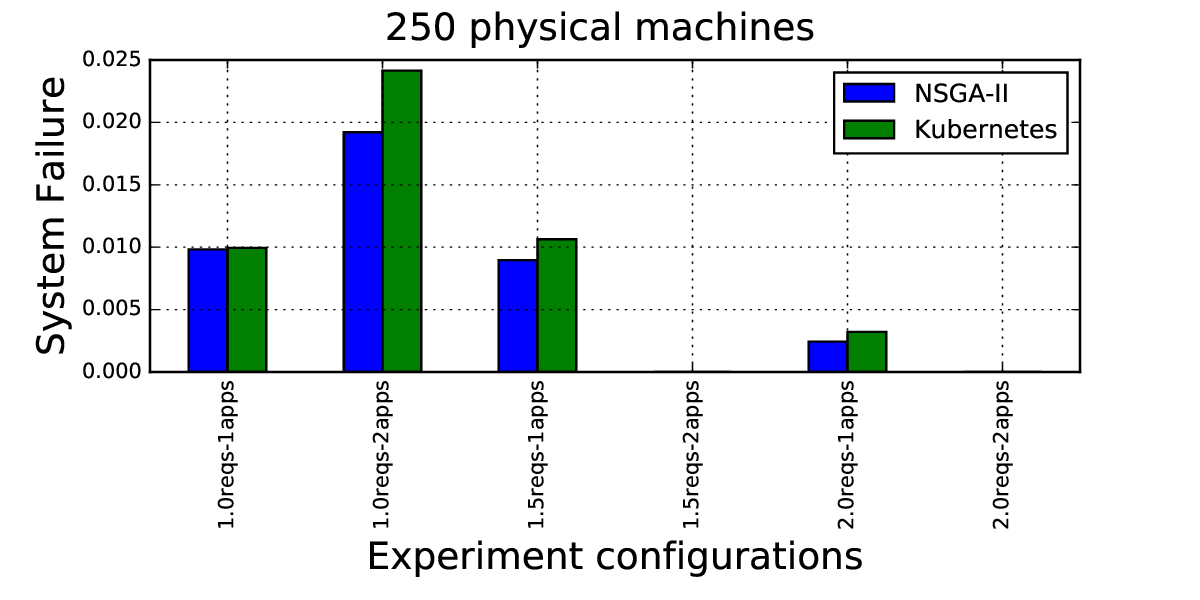}
	\includegraphics[width=0.48\textwidth]{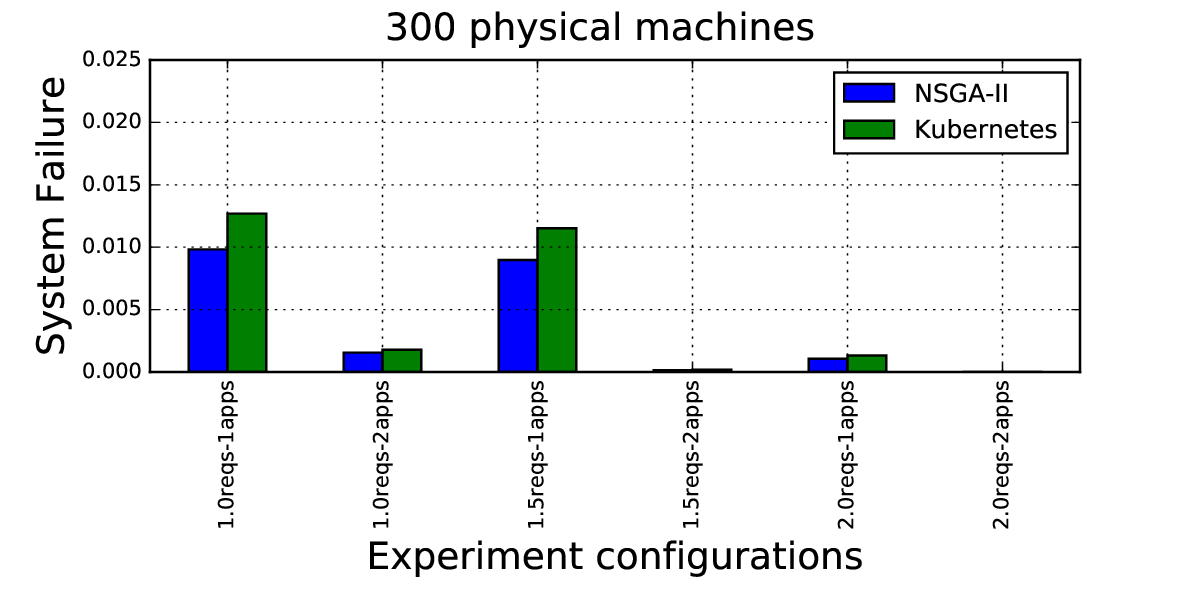}\\
	\includegraphics[width=0.48\textwidth]{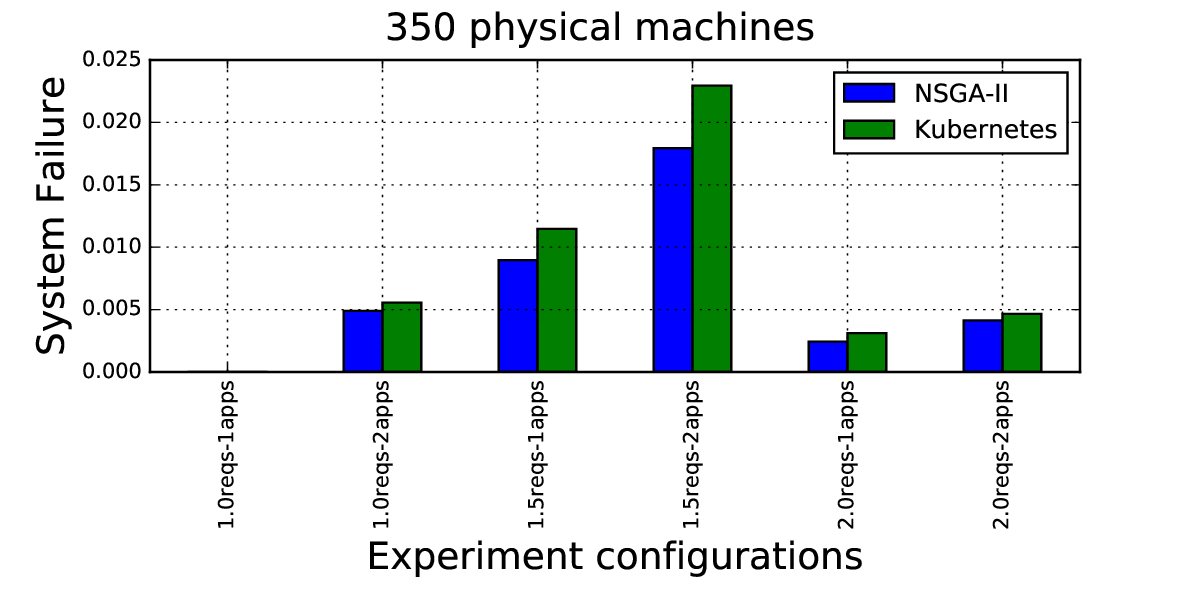}
	\includegraphics[width=0.48\textwidth]{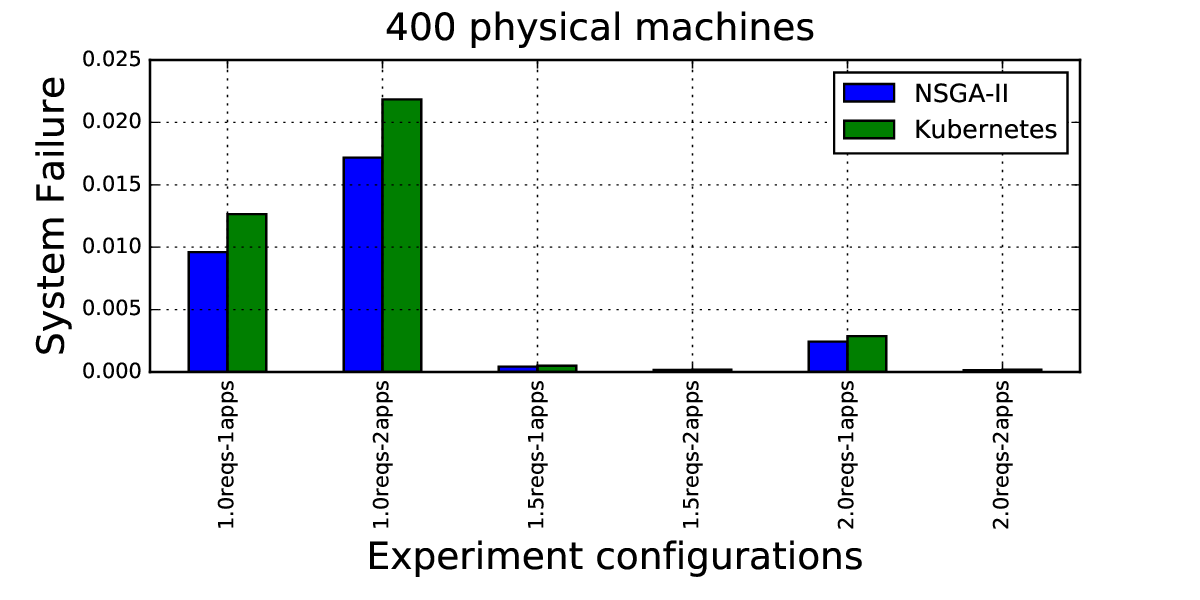}
	\caption{Comparative results for System Failure optimization}
	\label{fig:comparativeresultsrel}       % Give a unique label
\end{figure*}

\begin{figure*}
	\includegraphics[width=0.48\textwidth]{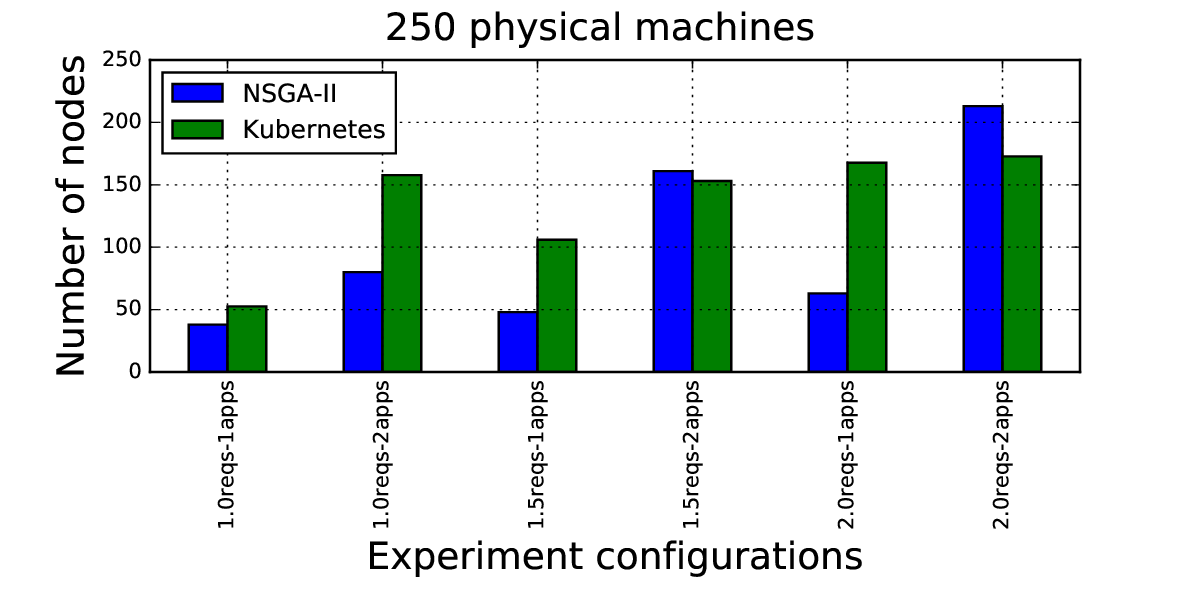}
	\includegraphics[width=0.48\textwidth]{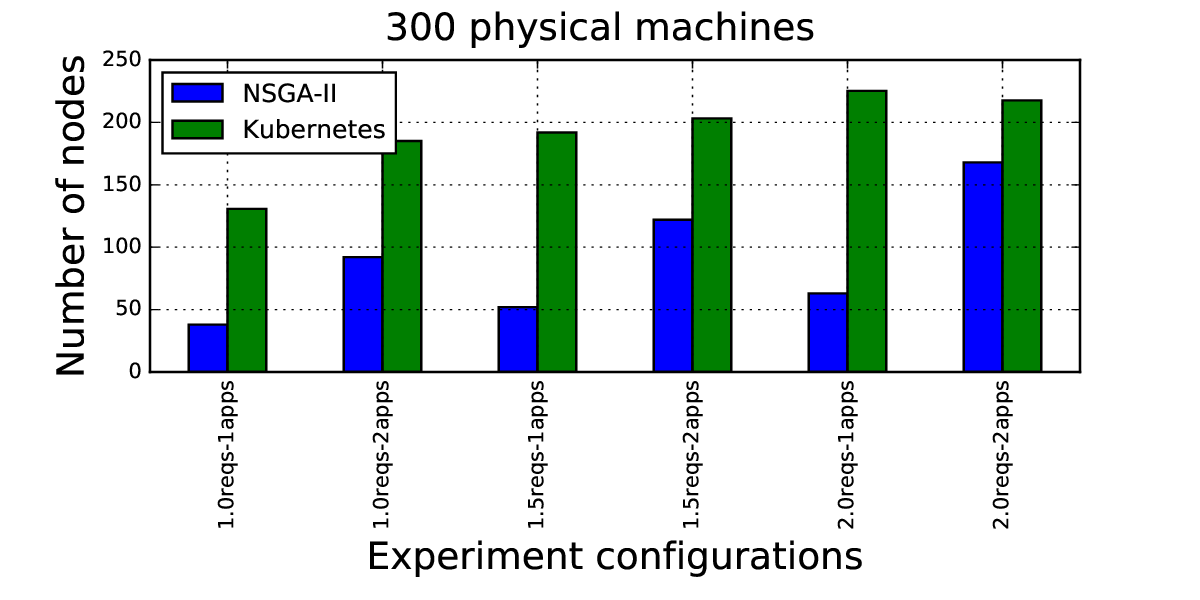}\\
	\includegraphics[width=0.48\textwidth]{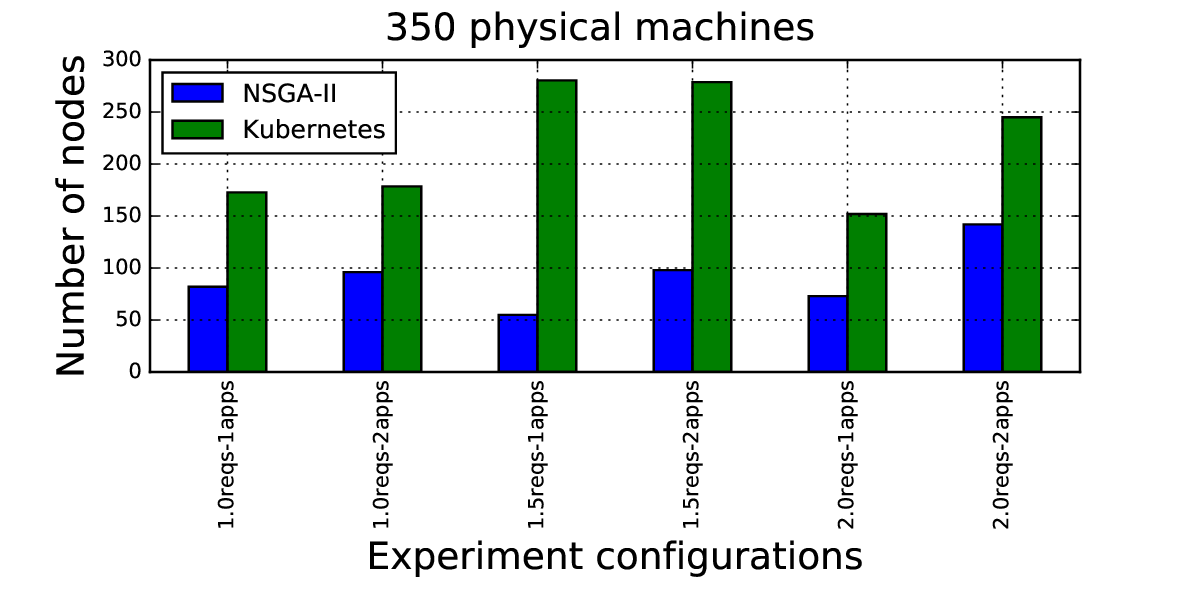}
	\includegraphics[width=0.48\textwidth]{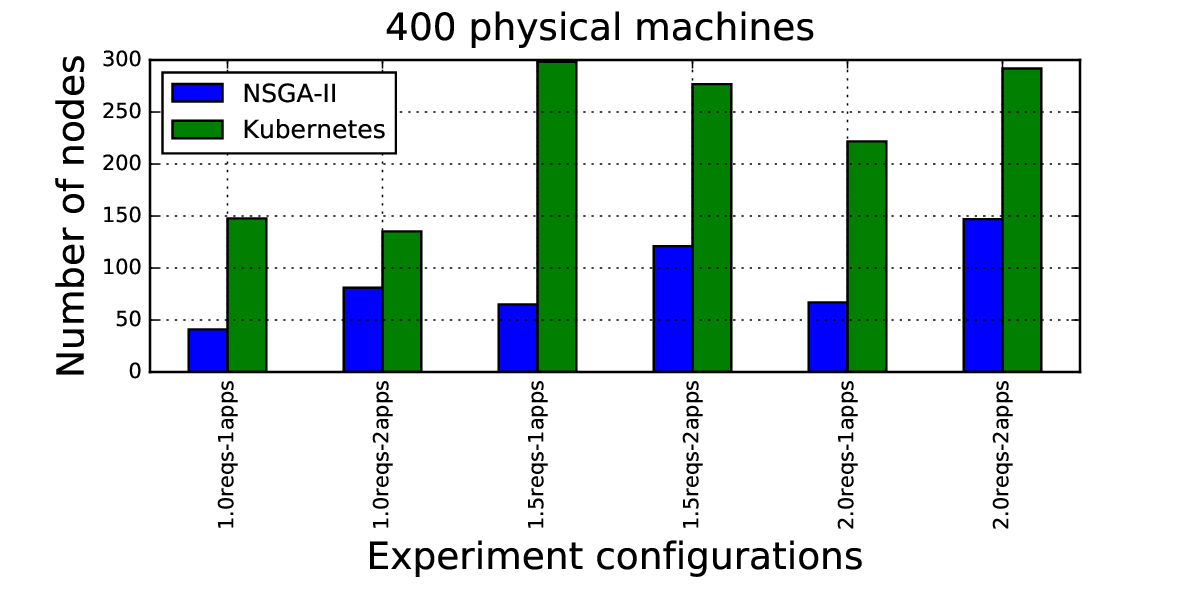}
	\caption{Comparative results for System Failure optimization}
	\label{fig:comparativeresultsnode}       % Give a unique label
\end{figure*}

In the second case, the analysis of final optimization results, we present the results for all the 24 experiment configurations. These results are plotted separately for each optimization function and for the total number of used physical machines: Figure~\ref{fig:comparativeresultsnet} for the Total Network Distance; Figure~\ref{fig:comparativeresultsthr} for Threshold Distance; Figure~\ref{fig:comparativeresultsclus} for the Cluster Balanced User; Figure~\ref{fig:comparativeresultsrel} for the System Failure; and Figure~\ref{fig:comparativeresultsnode} for the number of nodes with allocated containers. To represent a specific value for an experiment and an objective function, one solution among the solutions included in the Pareto front needs to be selected. We used again the criteria based on selecting the solution with the smaller SOV fitness value as we have previously explained.  The solutions using our approach, labeled as \textit{NSGA-II}, are compared with the solutions obtained using the allocation policies of Kubernetes, labeled as \textit{Kubernetes}.

\section{Discussion}

The analysis of the results, as in the case of their presentation in the previous section, is done from two points of views: the analysis of evolution of the objectives and the analysis of final optimized results.

\subsection{Evolution of the objectives}

In general terms, it is observed that the values for the objectives functions are established at the end of the execution of the optimization algorithm. This stabilization is even achieved before generation number 100 (Figures~\ref{fig:objvalues300-10-1}, \ref{fig:objvalues350-15-2} and~\ref{fig:objvalues400-20-1}). 

The \textit{min} series shows a very fast stabilization for the System Failure, Threshold Distance and Network Distance objectives. The \textit{min} series for Cluster Balanced is stabilized fairly later. In the case of the \textit{mean} series, it is observed that the values of the objectives are increased for some of them and decreased for other ones before they meet a stabilization. The reason is that the dominant solutions show equilibrated balances between all the objectives, thus some of them need to be deteriorated to allow the other objectives to be improved. 

We want to highlight the behavior of the representative solution of the Pareto front, the one with the smaller SOV value. In all the three cases, the value of the solution \textit{minSOV} is better than the values for the \textit{mean} series for some of the objectives and is worse for other objectives. For instance, in Figure~\ref{fig:objvalues400-20-1} the \textit{minSOV} solution is better for the System Failure, the Threshold Distance and the Network Distance, but worse for the Cluster Balanced. This represents  again the fact that for improving one of the objectives, it is necessary to degrade other objectives. It is important to remember that the values of the \textit{min} series for different objectives belong to different solutions and they represent the lowest value obtained considering all the solutions of our approach.

Finally, in relation with the number of physical machines with allocated containers (Figures~\ref{fig:sysvalues300-10-1}, \ref{fig:sysvalues350-15-2} and~\ref{fig:sysvalues400-20-1}), it is observed that the number of machines in use is smaller than the available ones for all the configurations. This is also a benefit, because other optimizations as energy consumption reduction can be achieve by switching off the machines with no allocated containers. Furthermore, the number of microservices represented in the figures indicates that the scale level for each container is between 18 and 25 for the \textit{mean} and between 2 and 7 for the \textit{min}, considering that there are 14 microservices for each application. This high level of scalability is explained because it benefits the reduction of the fail rate of the system, but it does not necessarily damage the other objectives, except for the Threshold Distance. A future improvement of the optimization process could be to include some type of constraint to avoid solutions with too high levels of scalability. 

Figures~\ref{fig:scatternetclus} and~\ref{fig:scatterfailthr} show how the initial number of solutions in the Pareto front is very small and how it increases as new generations are considered. We want to highlight how just in 30 generations the solutions in the Pareto front are shifted from random positions to the positions corresponding to the solution search space. From this generation, the solution space is not shifted but the number of solutions in the solution space is increased. This can be seen very clearly in  Figure~\ref{fig:scatterfailthr}, where due to the uniform distribution of the containers in the initial population, all the solutions in the initial Pareto front are concentrated in only one point of Threshold Distance and System Failure. In just 30 generations, the Pareto front is covering all the solution space, and in the following generations the solution space is populated with a bigger diversity of solutions. By this, the diversity of solutions with the same quality is increased and the covered solution space is bigger as new generations are obtained. 

We want to highlight that the solution search spaces of three of the objectives are well covered. But in the case of the System Failure objective, the solutions are not covering all the possible values (Figure~\ref{fig:scatterfailthr}). To guarantee a better distribution of the solutions across the solution search space in this objective, additional mutation operators should be considered.

In general terms, we can state that the use of a genetic approach based on NSGA-II is suitable to find optimized solutions with a reasonable number of generation (100) and a reasonable size of the population (200).

\subsection{Final optimized results}

In a second step, the final optimized results from all the experiment configurations are analyzed. The results for our approach correspond to the optimization values for the solution with a smaller SOV (Figures~\ref{fig:comparativeresultsnet}, \ref{fig:comparativeresultsthr}, \ref{fig:comparativeresultsclus} and~\ref{fig:comparativeresultsrel}). In general terms, we can observe that for the objectives Network Distance and Threshold Distance, the cases with the worst (higher) optimization values correspond to the ones with a configuration of 2 applications: as the number of microservices is increased, these two objectives values are worse. This behavior is also observed in the case of increasing the workload of the system (number or user application requests). On the contrary, the Cluster Balanced Use and the System Failure do no show a clear trend as the workload and number of applications is varied. Finally, the number of physical machines (Figure~\ref{fig:comparativeresultsnode}) seems to be influenced in the same way that the two first objectives functions: the total number of physical nodes is increased as the workload or the number of applications increase.

In relation with the results obtained with the Kubernetes approach, our solution showed better optimizations for all the objectives. It is important to highlight that these improvements were obtained in spite of using a smaller number of physical machines. This also supposed a collateral benefit because the machines with none allocated container could be switched off obtaining greater reduction in the power consumption of the cluster than in the case of Kubernetes.

\section{Conclusion}

In this article, we have addressed the problem of container resource allocation and elasticity (self-scaling of container replicas). A four-objectives optimization has been proposed based on a uniform distribution of the workload along the physical machines (Cluster Balanced Use), a tight distribution of the microservices workload along their container replicas (Threshold Distance), a reduction of the network overheads (Network Distance) and a better reliability (System Failure).

We have applied the NSGA-II, a genetic algorithm, to the problem of container allocation in cloud. Several experiments configurations, varying the number of physical machines and the workload of the system, have been tested in order to study the benefits of this algorithm. The results have shown that our approach is a suitable solution to address this problem and it found optimized solutions with reasonable number of generations (100) and reasonable size of the population (200).

The results have been compared with the obtained from the implementation of the Kubernetes' allocation policies. Our approach has shown a better behavior by obtaining a better optimization of the four objectives. Moreover, our solution uses a smaller number of physical machines.

As future work, we plan to study the results of our allocation algorithm in a real Cloud Container Cluster. Additionally, other optimization objectives can been included. Finally, container live-migration cost could be studied as the criteria to select solutions from the Pareto front, instead of using a criteria based on a single scalar value of the multi-objectives.

\bibliographystyle{spmpsci}      % mathematics and physical sciences
\bibliography{bibliography}   % name your BibTeX data base

\end{document}